\documentclass[pra, onecolumn, superscriptaddress, nofootinbib]{revtex4-2}
\usepackage[a4paper, left=1in, right=1in, top=1in, bottom=1in]{geometry}

\usepackage{cmap} 
\usepackage[utf8]{inputenc}
\usepackage[english]{babel}
\usepackage[T1]{fontenc}
\usepackage{booktabs}
\usepackage{appendix}
\usepackage{braket}
\usepackage{url, amsfonts, tikz, physics, listings, xcolor, graphicx, float}
\usepackage[shortlabels]{enumitem}
\usepackage{dsfont}
\usepackage{amsmath, amssymb, amstext, amscd, amsthm, makeidx, graphicx, url, mathrsfs, mathtools, longdivision, polynom, bbm, complexity, multirow}
\usepackage{nicefrac}
\usepackage[linesnumbered,ruled,vlined]{algorithm2e}
\usepackage{xcolor}
\usepackage[normalem]{ulem}
\usepackage{booktabs}

\definecolor{blueviolet}{rgb}{0.2, 0.2, 0.6}
\definecolor{webgreen}{rgb}{0,.5,0}
\definecolor{webbrown}{rgb}{.6,0,0}
\usepackage[pdftex,
  bookmarks=false,
  colorlinks=true, 
  urlcolor=webbrown,
  linkcolor=blueviolet, 
  citecolor=webgreen,
  pdfstartpage=1,
  pdfstartview={FitH},  
  bookmarksopen=false
  ]{hyperref}
\usepackage{cleveref}
\crefname{equation}{Eq.}{Eqs.}
\Crefname{equation}{Eq.}{Eqs.}

\makeatletter
\newcommand\RedeclareMathOperator{%
  \@ifstar{\def\rmo@s{m}\rmo@redeclare}{\def\rmo@s{o}\rmo@redeclare}%
}

\newcommand\rmo@redeclare[2]{%
  \begingroup \escapechar\m@ne\xdef\@gtempa{{\string#1}}\endgroup
  \expandafter\@ifundefined\@gtempa
     {\@latex@error{\noexpand#1undefined}\@ehc}%
     \relax
  \expandafter\rmo@declmathop\rmo@s{#1}{#2}}
\newcommand\rmo@declmathop[3]{%
  \DeclareRobustCommand{#2}{\qopname\newmcodes@#1{#3}}%
}
\@onlypreamble\RedeclareMathOperator
\makeatother

\allowdisplaybreaks

\RedeclareMathOperator*{\E}{{\mathbb{E}}}

\makeatletter
\DeclareFontFamily{OMX}{MnSymbolE}{}
\DeclareSymbolFont{MnLargeSymbols}{OMX}{MnSymbolE}{m}{n}
\SetSymbolFont{MnLargeSymbols}{bold}{OMX}{MnSymbolE}{b}{n}
\DeclareFontShape{OMX}{MnSymbolE}{m}{n}{
    <-6>  MnSymbolE5
   <6-7>  MnSymbolE6
   <7-8>  MnSymbolE7
   <8-9>  MnSymbolE8
   <9-10> MnSymbolE9
  <10-12> MnSymbolE10
  <12->   MnSymbolE12
}{}
\DeclareFontShape{OMX}{MnSymbolE}{b}{n}{
    <-6>  MnSymbolE-Bold5
   <6-7>  MnSymbolE-Bold6
   <7-8>  MnSymbolE-Bold7
   <8-9>  MnSymbolE-Bold8
   <9-10> MnSymbolE-Bold9
  <10-12> MnSymbolE-Bold10
  <12->   MnSymbolE-Bold12
}{}

\let\llangle\@undefined
\let\rrangle\@undefined
\DeclareMathDelimiter{\llangle}{\mathopen}%
                     {MnLargeSymbols}{'164}{MnLargeSymbols}{'164}
\DeclareMathDelimiter{\rrangle}{\mathclose}%
                     {MnLargeSymbols}{'171}{MnLargeSymbols}{'171}
\makeatother

\newtheorem{theorem}{Theorem}

\newtheorem{lemma}{Lemma}
\newtheorem{corollary}{Corollary}
\newtheorem{definition}{Definition}

\newtheorem{problem}{Problem}

\begin{document}
\title{Heisenberg-limited Hamiltonian learning without short-time control}

\author{Myeongjin Shin}
\email{hanwoolmj@kaist.ac.kr}
\affiliation{School of Computational Sciences, Korea Advanced Institute of Science and Technology, Daejeon 34141, Korea}

\author{Junseo Lee}
\email{junseolee@fas.harvard.edu}
\affiliation{Institute of Computer Technology, Seoul National University, Seoul 08826, Korea}

\author{Changhun Oh}
\email{changhun0218@gmail.com}
\affiliation{Department of Physics, Korea Advanced Institute of Science and Technology,
Daejeon 34141, Korea}

\date{\today}

\begin{abstract}
Characterizing quantum systems by learning their underlying Hamiltonians is a central task in quantum information science. While recent algorithmic advances have achieved near-optimal efficiency in this task, they critically rely on accessing arbitrarily short-time dynamics. This reliance poses severe experimental challenges due to finite control bandwidth and transient pulse errors. In this work, we demonstrate that Heisenberg-limited Hamiltonian learning can be achieved without short-time control. 
We introduce a framework in which every query to the unknown dynamics has duration at least a prescribed minimum time \(T\), and show that this restriction does not preclude Heisenberg-limited scaling. 
The key ingredient is a method for emulating the continuous quantum control required by iterative learning algorithms using only such lower-bounded evolution times. 
This reduces the learning task to sparse pure-state tomography.
Notably, for logarithmically sparse Hamiltonians, our algorithm achieves the information-theoretically optimal $1/\varepsilon$ scaling in total evolution time for \emph{any} arbitrary constant minimum evolution time $T$. For many-body (polynomially sparse) systems, we uncover a rigorous quantitative tradeoff, showing that the minimum required evolution time can be significantly relaxed from the standard limit at a polynomial cost in total evolution time. Our results affirmatively resolve a prominent open problem in the field and reveal that high-bandwidth, ultra-short pulses are not fundamentally necessary for optimal quantum learning.
\end{abstract}

\maketitle

\section{Introduction}

Understanding how interactions govern physical systems is a central goal of physics. 
In quantum systems, these interactions are encoded in the Hamiltonian, which dictates the system's dynamical behavior. 
Precisely characterizing this Hamiltonian from observed dynamics has become a fundamental problem of broad interest, with applications ranging from quantum metrology to quantum device benchmarking and the verification of analog quantum simulators~\cite{wang2017experimental, pang2014quantum, shaffer2021practical, gu2024practical}. 
A central question is therefore whether one can efficiently learn the interactions governing an unknown quantum system directly from its time dynamics, and under what control assumptions this is possible.

The problem of learning an unknown Hamiltonian from time evolution has been extensively studied~\cite{huang2023learning, yu2023robust, ma2024learning, hu2025ansatz, abbas2025nearly, gu2024practical, li2024heisenberg, zhao2025learning, da2011practical, wiebe2014hamiltonian, dutkiewicz2024advantage, wang2017experimental, pang2014quantum, shaffer2021practical, zubida2021optimal, haah2022optimal}. 
A standard theoretical framework assumes that the Hamiltonian is $m$-sparse in the Pauli basis,
\begin{equation}
    H = \sum_{x=1}^m \alpha_x P_x,    
\end{equation}
where $\{P_x\}$ are Pauli operators and $\{\alpha_x\}$ are unknown nonzero coefficients. 
Given oracle access to the time-evolution operator $e^{-iHt}$ at chosen time durations $\{t_i\}_{i=1}^{\mathcal{N}}$, the goal is to output an estimate $\widetilde{H}$ that approximates $H$ to the desired accuracy. 
To evaluate the efficiency of learning algorithms in this setting, two natural complexity measures are commonly considered:
\begin{equation}
    \text{total evolution time}: \; t_{\mathrm{tot}} = \sum_{i=1}^\mathcal{N} t_i,
    \qquad
    \text{minimum evolution time}: \; t_{\min} = \min_{i \in [\mathcal{N}]} t_i,
\end{equation}
where $\mathcal N$ denotes the number of time-evolution queries used by the algorithm.
While $t_{\mathrm{tot}}$ captures the overall dynamical cost of the protocol, $t_{\min}$ quantifies the shortest timescale that the learner must physically access, and is therefore the quantity most directly tied to experimental control constraints. 
Our focus in this work is precisely to understand whether Heisenberg-limited Hamiltonian learning remains possible when $t_{\min}$ is lower bounded by an arbitrary fixed constant $T>0$.

Recent works have achieved strong guarantees for sparse Hamiltonian learning, including nearly optimal dependence on the target precision in terms of total evolution time~\cite{bakshi2024structure, hu2025ansatz, abbas2025nearly}. 
A key requirement of these state-of-the-art approaches is access to short-time discrete control, namely, the ability to access sufficiently fine-grained time evolution, including arbitrarily short time dynamics.
This capability is crucial in iterative refinement schemes: given a current estimate $H_j$ of the unknown Hamiltonian $H$, the learner seeks to isolate the remaining unknown interactions by simulating evolution under the residual Hamiltonian $\Delta H_j = H - H_j$. 
This requirement arises from the need to implement continuous quantum control $e^{-i\Delta H_j t}$ through multiple applications of discrete control, as learning is fundamentally limited to the standard quantum limit in the absence of discrete control~\cite{dutkiewicz2024advantage}.

To synthesize this target dynamics, standard approaches combine queries to the physical evolution \(e^{-iHt}\) with simulation of the current estimate \(H_j\), and rely on Trotterization~\cite{childs2021theory, blanes2025concise, berry2007efficient}:
\begin{equation}
    e^{-i\Delta H_j t} \simeq \left(e^{-iHt/N}e^{iH_jt/N}\right)^N.
\end{equation}
For this product formula to accurately approximate the target evolution, the number of segments $N$ must be sufficiently large. 
As a result, the elementary query time $t/N$ must become correspondingly small, so access to vanishingly short-time dynamics is built directly into the simulation primitive. 
In particular, previous algorithms require the minimum evolution time to scale as $t_{\min} = \Theta(1/m)$~\cite{bakshi2024structure} or $t_{\min} = \Theta(\sqrt\varepsilon)$~\cite{abbas2025nearly, hu2025ansatz}, so the accessible time window must shrink as the sparsity $m$ grows or the target accuracy $\varepsilon$ is tightened.

Implementing such extremely short-time evolutions---especially when they must be repeatedly interleaved with additional quantum control operations---poses a significant challenge for physical hardware~\cite{oh2002errors}. 
In realistic platforms, control pulses have finite rise and fall times, so when the intended evolution window becomes too short, transient distortions and switching errors can become comparable to, or even dominate, the desired Hamiltonian dynamics~\cite{gustavsson2013improving,jerger2019situ,rol2020timedomain}.
By contrast, over longer continuous evolutions, these switching imperfections are amortized and typically represent only a small fraction of the total dynamics. 
This asymmetry becomes particularly problematic for Trotterization-based protocols, where the target evolution is decomposed into many short segments and the switching overhead is paid anew at every step. In practice, the required pulse sequencing is therefore constrained not only by timing precision but also by finite control bandwidth and by dead times between successive control operations. 
These overheads quickly consume the available coherence budget and can substantially degrade the fidelity of the overall learning procedure. 
Together, they severely limit the practical viability of Hamiltonian-learning frameworks that rely on ultra-short-time dynamics.


These physical constraints strongly motivate studying Hamiltonian learning in a fundamentally different regime: one in which \emph{short-time evolution} is not accessible. 
Concretely, a setting where every available evolution time is lower bounded by a constant $T$, independent of both the sparsity $m$ and the target accuracy $\varepsilon$. 
This leads to the following central question:
\begin{center}
    \textit{Can one achieve Heisenberg-limited Hamiltonian learning without using short-time evolution?}
\end{center}


\begin{figure}
    \centering
    \includegraphics[width=\linewidth]{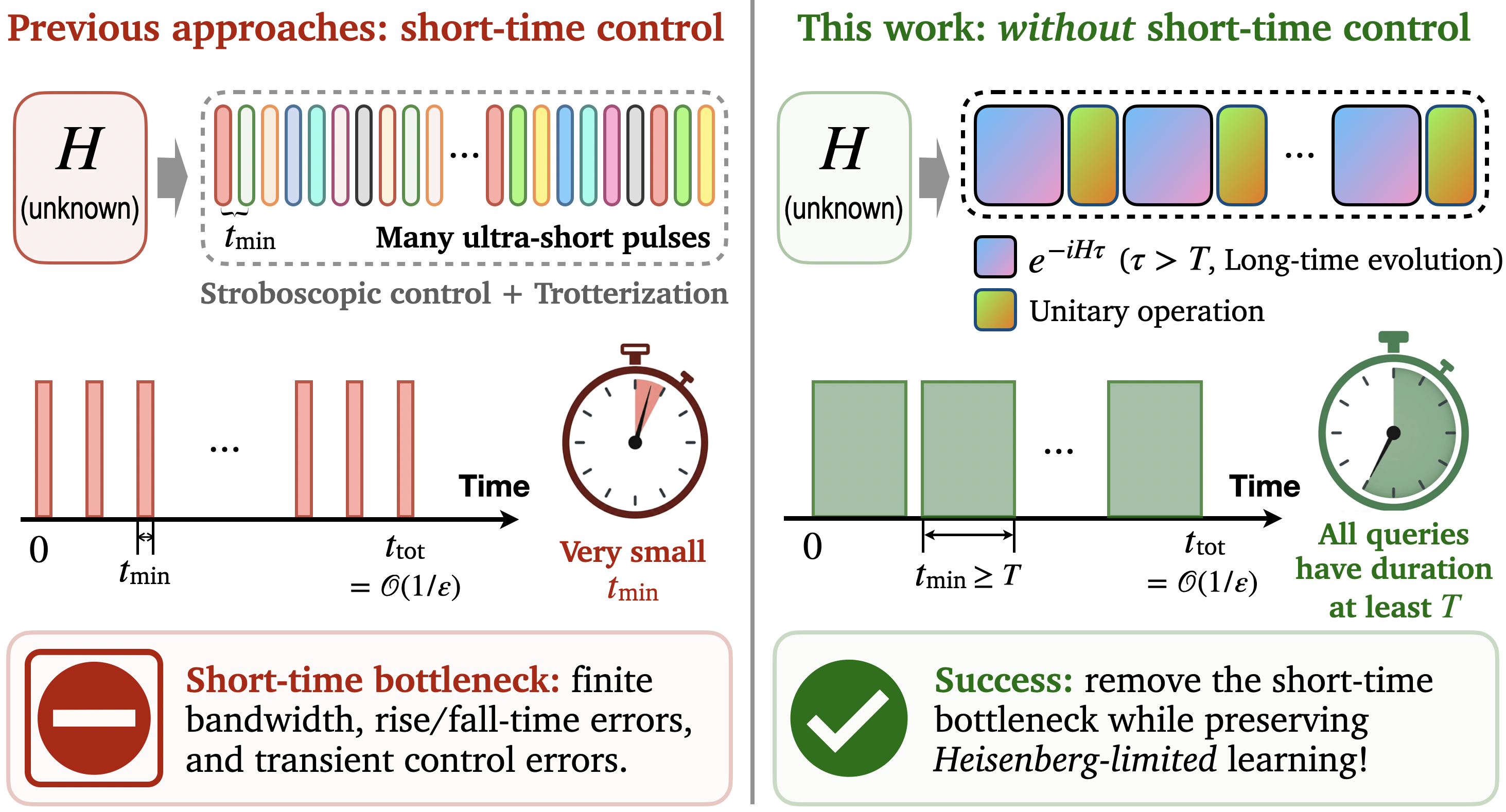}
    \caption{
    Comparison between previous short-time control approaches and our framework. 
    Previous Heisenberg-limited algorithms rely on stroboscopic control and Trotterization, requiring many ultra-short evolution segments with small \(t_{\min}\). 
    Our framework replaces these short-time queries by long-time evolutions interleaved with learned auxiliary operations, ensuring that every query to the unknown Hamiltonian has duration at least \(T\). 
    Thus, the short-time bottleneck is removed while Heisenberg-limited scaling in total evolution time is retained.
    }
    \label{fig:overview}
\end{figure}

\medskip

\noindent\textbf{Our contribution and technical overview.}
In this work, we show that Heisenberg-limited Hamiltonian learning can be achieved without access to arbitrarily short-time dynamics. 
For logarithmically sparse Hamiltonians, every query to the unknown evolution can have duration at least an arbitrary fixed constant \(T>0\), while the total evolution time still scales as \(1/\varepsilon\) in the target accuracy and remains efficient in the number of qubits.
For polynomially sparse Hamiltonians, we obtain a quantitative tradeoff between the minimum accessible evolution time and the total evolution time. In particular, one can achieve a constant minimum evolution time independent of \(m\) at the cost of a quasi-polynomial overhead in the total evolution time, while still retaining Heisenberg-limited scaling in \(\varepsilon\).

The central difficulty is that existing Heisenberg-limited algorithms rely on residual dynamics. 
Given a current estimate \(H_j\), one would like to isolate the remaining Hamiltonian \(\Delta H_j=H-H_j\) by implementing \(e^{-i\Delta H_j t}\). 
Standard product-formula implementations do this through short steps such as
\begin{equation}
    \left(e^{-iHt/N}e^{iH_jt/N}\right)^N,
\end{equation}
which directly requires access to \(e^{-iHt/N}\). 
Our main observation is that each such short step can be rewritten as a long-time step plus a correction:
\begin{equation}
    e^{-iH\tau}e^{iH_j\tau}
    =
    e^{-iH(T+\tau)}C_j e^{iH_j(T+\tau)},
    \qquad
    C_j=e^{iHT}e^{-iH_jT}.
\end{equation}
Thus, all queries to the unknown Hamiltonian have duration at least \(T\).

The correction \(C_j\) is not directly accessible because it contains backward evolution under the unknown Hamiltonian. 
However, its adjoint \(C_j^\dagger=e^{iH_jT}e^{-iHT}\) is accessible using only forward long-time evolution under \(H\) and simulation of the known Hamiltonian \(H_j\). 
Our strategy is therefore to learn a Hermitian generator $W$ of \(C_j^\dagger=e^{-iW}\), and use it to implement the inverse correction needed in the long-time product formula. 
The key technical point is that this auxiliary generator has bounded norm and sparsity, or quasi-sparsity, to be learned and simulated efficiently.

Once residual dynamics can be emulated, the remaining task is coefficient extraction. 
We reduce this step to sparse pure-state tomography: applying the residual dynamics \(e^{-i\Delta H_j t}\) to one half of a maximally entangled state encodes the Pauli coefficients of \(\Delta H_j\) as first-order amplitudes. 
By recovering these amplitudes and applying truncation to enforce bounded norm and sparsity, we obtain a refined estimate \(H_{j+1}\). 

Iterating this refinement yields the main guarantees of the paper. 
For logarithmically sparse Hamiltonians, our algorithm allows any fixed minimum evolution time \(T>0\) while retaining Heisenberg-limited scaling in the target accuracy. 
For polynomially sparse Hamiltonians, it gives an explicit tradeoff between the minimum accessible evolution time and the total evolution time. 
Thus, the need for ultra-short-time dynamics can be removed entirely in the logarithmically sparse regime and substantially relaxed in the many-body regime. To the best of our knowledge, this is the first work to establish a quantitative tradeoff between the minimum accessible evolution time and the total evolution time in Hamiltonian learning. 
Our results resolve an open problem posed by Bakshi, Liu, Moitra, and Tang~\cite{bakshi2024structure}. 
More broadly, we suggest a conceptual shift in quantum learning from dynamics: access to ultra-short evolution times is not a fundamental requirement for information-theoretically optimal learning.
We now formalize this learning model that avoids short-time evolution and state the corresponding guarantees.

\section{Problem setup and main results}

We now formalize the long-time Hamiltonian learning problem and state the main results. 
The purpose of this section is to specify the Hamiltonian model, the allowed oracle access, and the resource measures used throughout the paper.

\medskip
\noindent\textbf{Hamiltonian model.}
Let $H \in \mathbb{C}^{2^n \times 2^n}$ be an $n$-qubit, $m$-sparse, traceless Hamiltonian of the form
\begin{equation}\label{eq:ham-def}
    H = \sum_{x=1}^m \alpha_x P_x,
\end{equation}
where $P_x \in \{\mathbb{I},X,Y,Z\}^{\otimes n}$ are $n$-qubit Pauli operators and $\alpha_x \in \mathbb{R}$ are unknown nonzero coefficients.
The sparsity parameter $m$ reflects the underlying physical structure of the system: the regime $m = \mathcal{O}(\log n)$ characterizes highly structured systems where a few key interactions dominate the dynamics, whereas $m = \text{poly}(n)$ encompasses the broader class of many-body Hamiltonians typically encountered in condensed matter and quantum information science.

\medskip
\noindent\textbf{Access model.}
We are given oracle access to the time-evolution unitary $U(t) = e^{-iHt}$, for any evolution time $t \ge T$, where $T>0$ is an arbitrarily large constant. That is, for each query the learner may choose any time $t \ge T$ and obtain access to the corresponding unitary $U(t)$.

\medskip
\noindent\textbf{Complexity measures.}
We evaluate the efficiency of a learning algorithm using the following quantities:
\begin{itemize}
    \item the \emph{total evolution time} $t_{\mathrm{tot}} = \sum_{i=1}^\mathcal{N} t_i$,
    
    \item the \emph{minimum evolution time} $t_{\min} = \min_{i \in [\mathcal{N}]} t_i$,
    
    \item the \emph{number of queries} $\mathcal{N}$.
\end{itemize}
Here $t_1,\dots,t_{\mathcal N}$ denote the evolution times used by the algorithm. In our setting, the constraint $t_{\min} \ge T$ prohibits access to arbitrarily short-time evolution.

\medskip
\noindent\textbf{Definitions.}
Let $H=\sum_{x=1}^m \alpha_x P_x$ be an $n$-qubit Hamiltonian with eigenvalues $\{\lambda_j\}_{j=1}^{2^n}$. We use the following notions of norm and sparsity:
\begin{itemize}
    \item \textbf{Matrix norms:}
    \begin{equation}
        \|H\|_\infty := \max_{j \in [2^n]} |\lambda_j|,
    \qquad
    \|H\|_F := \sqrt{\frac{1}{2^n}\sum_{j=1}^{2^n} \lambda_j^2}
    = \sqrt{\sum_x \alpha_x^2}.    
    \end{equation}
    Here $\|\cdot\|_\infty$ denotes the operator norm, and $\|\cdot\|_F$ denotes the normalized Frobenius norm.

    \item \textbf{Unitary distance:}
    \begin{equation}
    d(U,V) := \min_{\phi} \left\|U-e^{i\phi}V\right\|_F.
    \end{equation}
    Here $d(U,V)$ denotes the distance between unitaries up to a global phase.
    
    \item \textbf{(Pauli) coefficient norms:} for $p \ge 1$,
    \begin{equation}
        \|H\|_{\ell_p} := \left(\sum_x |\alpha_x|^p\right)^{1/p},
    \qquad
    \|H\|_{\ell_\infty} := \max_x |\alpha_x|.    
    \end{equation}

    \item \textbf{(Pauli) sparsity:}
    \begin{equation}
        \operatorname{supp}_P(H) := \left|\{x : \alpha_x \neq 0\}\right|.
    \end{equation}
    In particular, if $H$ is given as in \Cref{eq:ham-def}, then $\operatorname{supp}_P(H)=m$.

    \item \textbf{(Pauli) quasi-sparsity:} for $\varepsilon \ge 0$,
    \begin{equation}
        \operatorname{supp}_P(H,\varepsilon)
    := \min \left\{
    \operatorname{supp}_P(\widetilde{H}) :
    \|H-\widetilde{H}\|_F \le \varepsilon
    \right\}.    
    \end{equation}
\end{itemize}

\medskip
Combining the above ingredients, we arrive at the central problem studied in this work.

\begin{problem}[Hamiltonian learning with minimum evolution time $T$]\label{problem:longtimeevol}
Let $H$ be an $n$-qubit, $m$-sparse, traceless Hamiltonian of the form $H=\sum_{x=1}^m \alpha_x P_x$. Suppose we are given oracle access to the time-evolution unitary $U(t)=e^{-iHt}$, for any evolution time $t \ge T$, where $T>0$ is an arbitrarily large constant.
The goal is to output an estimate $\widetilde{H}$ such that $\|H-\widetilde{H}\|_{\ell_\infty} < \varepsilon$.
\end{problem}

We say that the problem admits an efficient learning algorithm if the total evolution time is $\mathrm{poly}(n)/\varepsilon$, the number of queries is $\mathrm{poly}(n, 1/\varepsilon)$, and the classical post-processing time is $\mathrm{poly}(n,1/\varepsilon)$.

This formalization captures a realistic experimental regime in which the learner is restricted to accessing only long-time dynamics, thereby avoiding transient control errors associated with ultra-short evolutions. The key challenge is to extract high-precision information—namely, the individual interaction coefficients \(\{\alpha_x\}\)—from such coarse, long-duration evolutions. The following results show that this is possible in logarithmically sparse systems and remains possible for non short regimes in $T$, with a quantitative tradeoff, for polynomially sparse systems.

\medskip

\noindent\textbf{Main results.}
We now state our main guarantees for Hamiltonian learning under the minimum-evolution-time constraint in \Cref{problem:longtimeevol}. 
The first result applies to logarithmically sparse Hamiltonians. 
In this regime, an exponential dependence on the sparsity \(m\) remains efficient in the system size, and the minimum accessible evolution time can be chosen as an arbitrary fixed constant.

\begin{theorem}[Learning logarithmically sparse Hamiltonians from arbitrary long-time evolution]
\label{thm:arb-long}
Let \(H\) be an \(n\)-qubit, \(m\)-sparse traceless Hamiltonian with \(\|H\|_{\infty} \le 1\).
There exists a learning algorithm that solves~\Cref{problem:longtimeevol}
with access to \(e^{-iHt}\) for all \(t \ge T\), using total evolution time
\begin{equation}
    t_{\mathrm{tot}} =
    \widetilde{\mathcal{O}}\left(
    \min\left\{
    \frac{4^m T^3}{\varepsilon},
    \frac{4^m T}{\varepsilon^2}
    \right\}
    \right),
\end{equation}
and minimum evolution time \(t_{\min}=T\), with high probability. 
In particular, when \(m=\mathcal{O}(\log n)\), this gives an efficient algorithm for any fixed constant \(T>0\), achieving Heisenberg-limited scaling in the target accuracy while never querying the unknown dynamics for times shorter than \(T\).
\end{theorem}

The overall complexity is given by the minimum of the Heisenberg-limited and standard quantum limit scalings. 
In the small-error regime, the Heisenberg-limited scaling $1/\varepsilon$ dominates. 
In the complementary regime, the $1/\varepsilon^2$ scaling applies; however, since the error $\varepsilon$ is bounded below by a constant independent of the desired accuracy, the total evolution time in this regime is effectively a constant in terms of $\varepsilon$. 
Consequently, combining both regimes, the overall complexity is governed by the Heisenberg-limited scaling.

The next result extends the framework to many-body Hamiltonians with polynomial sparsity. 
Here the minimum accessible evolution time can still be made substantially larger than the \(1/m\)-scale required by standard product-formula approaches, at the cost of an increased polynomial dependence on \(m\).

\begin{theorem}[Learning many-body Hamiltonians without short-time evolution]
\label{thm:many-long}
Let \(H\) be an \(n\)-qubit, \(m\)-sparse traceless Hamiltonian with \(\|H\|_{\infty} \le 1\).
There exists a learning algorithm that solves~\Cref{problem:longtimeevol}
with access to \(e^{-iHt}\) for all \(t \ge T\), using total evolution time
\begin{equation}
t_{\mathrm{tot}}
=
\widetilde{\mathcal{O}}\left(
\min\left\{
\frac{m^{K+2}T}{\varepsilon},
\frac{m^KT}{\varepsilon^2}
\right\}
\right),
\end{equation}
provided that the minimum evolution time satisfies \(t_{\min}=T=\Theta(m^{-1/K})\), where \(K\in\mathbb{N}\). 
The algorithm succeeds with high probability.

Moreover, choosing \(K=\Theta(\log m)\) makes the minimum evolution time constant, \(t_{\min}=\Theta(1)\), at the expense of quasi-polynomial total evolution time,
\begin{equation}
t_{\mathrm{tot}}
=
\widetilde{\mathcal{O}}\left(
\min\left\{
\frac{m^{2+\log m}}{\varepsilon},
\frac{m^{\log m}}{\varepsilon^2}
\right\}
\right).
\end{equation}
\end{theorem}

These results show that the minimum accessible evolution time and the total evolution time are tradable resources in Hamiltonian learning. 
For logarithmically sparse Hamiltonians, the short-time requirement can be removed entirely while retaining Heisenberg-limited scaling. 
For polynomially sparse Hamiltonians, our bounds give an explicit tradeoff between increasing \(t_{\min}\) and increasing \(t_{\mathrm{tot}}\). In particular, they allow constant \(t_{\min}\) at the expense of quasi-polynomial total evolution time.
The main remaining open question is whether polynomially sparse Hamiltonians can be learned with polynomial total evolution time while allowing \(t_{\min}\) to be an arbitrary fixed constant independent of \(m\). 
We next describe the learning architecture that proves \Cref{thm:arb-long,thm:many-long}.


\section{Algorithm overview}
We now describe how the main algorithm is organized, which is illustrated in Fig.~\ref{fig:algorithm}.
The protocol follows an iterative refinement scheme. 
Starting from \(H_0=0\), the learner maintains an estimate \(H_j\) with
\[
    \|H-H_j\|_{\ell_\infty}\le \eta_j,
\]
and aims to produce \(H_{j+1}\) with \(\eta_{j+1}=\eta_j/2\). 
Thus each round reduces to learning the residual Hamiltonian
\[
    \Delta H_j=H-H_j
\]
to accuracy comparable to \(\eta_j\).

Each refinement round has two conceptual steps. 
First, the learner emulates the residual evolution \(e^{-i\Delta H_j t_j}\) using only queries to the unknown dynamics \(e^{-iHt}\) with \(t\ge T\), together with simulations of known Hamiltonians. 
Second, the learner converts this residual evolution into estimates of the Pauli coefficients of \(\Delta H_j\) by applying it to one half of a maximally entangled state and performing sparse pure-state tomography.

Schematically, one round takes the form
\[
H_j
\;\xrightarrow{\text{long-time emulation}}
\widetilde U_{\mathrm{res}}^{(j)} \approx e^{-i(H-H_j)t_j}
\;\xrightarrow{\text{sparse tomography}}
\widetilde{\Delta H}_j
\;\xrightarrow{\text{truncate and update}}
H_{j+1}.
\]
The truncation step keeps the estimate \(m\)-sparse and norm bounded, ensuring that the next round satisfies the same structural promises.

The key new ingredient is the first step. 
Instead of implementing residual dynamics through short product-formula steps, we rewrite each short step as a long-time query plus a correction unitary. 
This correction is implemented by learning a structured generator \(W_j\) from the accessible adjoint dynamics \(e^{iH_jT}e^{-iHT}\). 
The norm and sparsity, or quasi-sparsity, of \(W_j\) determine the final complexity.

The algorithm uses this Heisenberg-limited refinement once the current error is sufficiently small. 
For earlier rounds, where the residual error is large, we use a simpler standard-quantum-limit subroutine. 
The full protocol and the crossover between the two regimes are given in \Cref{sec:full-protocol}. 
We first develop the long-time emulation primitive, which is the main technical contribution.

\begin{figure}
    \centering
    \includegraphics[width=\linewidth]{algorithm.png}
    \caption{
    Schematic overview of our iterative learning protocol. 
    At iteration \(j\), the current estimate \(H_j\) defines a residual Hamiltonian \(\Delta H_j=H-H_j\). 
    We first learn an auxiliary Hamiltonian \(W_j\) using only long-time evolutions under \(H\), and then use \(W_j\) to emulate the residual-control primitive \(e^{-i\Delta H_j t}\). 
    This emulated evolution reduces coefficient extraction to sparse pure-state tomography. 
    After estimating \(\Delta H_j\), we update and truncate the estimate to obtain \(H_{j+1}\), preserving \(m\)-sparsity and bounded norm while halving the error.
    }
    \label{fig:algorithm}
\end{figure}

\section{Long-time emulation of residual dynamics}\label{sec:long-time-emulation}

We now construct the first ingredient of the learning algorithm: a long-time implementation of the residual evolution
\begin{equation}
    e^{-i\Delta H_j t}=e^{-i(H-H_j)t},
\end{equation}
given a current estimate \(H_j\). 
Throughout this section, \(H\) and \(H_j\) are \(m\)-sparse and satisfy \(\|H-H_j\|_{\ell_\infty}\le \varepsilon\), while the learner may query the unknown dynamics only for times \(\tau\ge T\).

This primitive is essential for isolating the remaining unknown coefficients during iterative refinement, but standard Hamiltonian-learning protocols implement it using short-time queries to the unknown dynamics. 
Our goal in this section is to construct the same primitive using exclusively long-time queries \(e^{-iH\tau}\) with \(\tau\ge T\). 

\medskip
\noindent
\textbf{Standard approach: Trotterization.}
The conventional method for approximating continuous quantum control relies on Trotterization~\cite{childs2021theory, blanes2025concise, berry2007efficient}:
\begin{equation}\label{eq:standard-trotter-bound}
\left\|\left(e^{-iHt/N}e^{iH_jt/N}\right)^N-e^{-i(H-H_j)t}\right\|_F
\le
\frac{t^2 \min\{\|H\|_{\infty},\|H_j\|_{\infty}\}}{N}\|H-H_j\|_F.
\end{equation}
This approximation simulates $e^{-i(H-H_0)t}$ by repeatedly applying the discrete control operation 
\begin{equation}
    e^{-iHt/N}e^{iH_jt/N}.    
\end{equation}
However, this approach intrinsically requires access to short-time evolution $e^{-iHt/N}$, which violates the long-time evolution constraint considered in this work. 

This is precisely the short-time bottleneck inherited by previous Heisenberg-limited Hamiltonian-learning algorithms. 
In an iterative learning scheme, one repeatedly constructs a current estimate \(H_j\) and then uses residual dynamics generated by \(H-H_j\) to refine the estimate. 
When this residual dynamics is implemented through the product formula above, every refinement step requires many short applications of the unknown evolution \(e^{-iHt/N}\), interleaved with simulations of the known estimate \(H_j\). 
Consequently, the minimum accessible evolution time becomes tied to the product-formula step size. 
As the desired accuracy improves, or as the sparsity-dependent control error becomes more stringent, the required step size must shrink. 
This is the mechanism behind the ultra-short-time requirements in existing approaches.

Our goal is to break this link. 
We show that each short-time product-formula step can be replaced by a long-time query to the unknown Hamiltonian, together with a correction unitary that is itself accessible through long-time dynamics.

\medskip
\noindent
\textbf{Our approach: Emulating short-time steps via long-time dynamics.}
To bypass the need for short-time evolution, we rewrite each elementary product-formula step as a long-time evolution interrupted by a fixed correction unitary. 
Define
\begin{equation}
    C_j = e^{iHT}e^{-iH_jT}.
\end{equation}
Then, for any $\tau \ge 0$,
\begin{equation}
\label{eq:long-time-step}
    e^{-iH\tau}e^{iH_j\tau}
    =
    e^{-iH(T+\tau)}C_je^{iH_j(T+\tau)} .
\end{equation}
Indeed, substituting the definition of \(C_j\) cancels the additional evolutions of duration \(T\). 
Setting \(\tau=t/N\), we obtain the exact rewriting
\begin{equation}
\label{eq:long-time-trotter}
\left(e^{-iHt/N}e^{iH_jt/N}\right)^N
=
\left(
e^{-iH(T+t/N)}C_je^{iH_j(T+t/N)}
\right)^N .
\end{equation}
Crucially, every query to the unknown evolution on the right-hand side has duration at least \(T\).
Thus, if the correction unitary \(C_j\) could be implemented, the usual product-formula approximation to \(e^{-i(H-H_j)t}\) would require no short-time evolution under \(H\).

At first sight, this rewriting only shifts the difficulty from short-time evolution to the implementation of the correction unitary \(C_j\). 
The useful viewpoint is to regard \(C_j\) not as a black-box unitary to be estimated directly, but as a correction to be implemented through a learned Hermitian generator.
That is, we seek a traceless Hermitian matrix \(W_j\) satisfying $d(C_j,e^{iW_j})=0.$
If such a generator \(W_j\) is known explicitly, then the missing correction \(C_j\) can be implemented by simulating the known Hamiltonian \(W_j\) to realize \(e^{iW_j}\).
In this sense, the problem of removing short-time queries is reduced to the problem of learning a suitable generator \(W_j\) for the correction unitary.

In the actual learning protocol, we only construct a finite-precision estimate \(\widetilde W_j\) of \(W_j\). 
Suppose that \(\widetilde W_j\) satisfies
\begin{equation}
    \left\|W_j-\widetilde W_j\right\|_F \le \frac{\varepsilon'}{2N}.
\end{equation}
Then replacing \(C_j=e^{iW_j}\) by \(e^{i\widetilde W_j}\) in the long-time product formula introduces at most \(\varepsilon'\) error over the \(N\) product-formula steps. 
By choosing \(N\) sufficiently large such that the right-hand side of \Cref{eq:standard-trotter-bound} is at most \(\varepsilon'/2\), and by learning \(W_j\) to the corresponding accuracy, the product-formula and replacement errors can be jointly bounded by \(\varepsilon'\):
\begin{equation}\label{eq:sim-con-quantum-control}
    d\!\left(e^{-i(H-H_j)t}, 
    \left(e^{-iH(T+t/N)}e^{i\widetilde{W_j}}e^{iH_j(T+t/N)}\right)^N
    \right)
    \le \frac{\varepsilon'}{2}+\frac{\varepsilon'}{2N} \cdot N = \varepsilon'.
\end{equation}
This approximate continuous quantum control primitive will be used as a subroutine in the main Hamiltonian learning algorithm (Algorithm~\ref{alg:main}).

The usefulness of this reduction depends on whether the generator \(W_j\) can be learned and simulated efficiently. 
The generator \(W_j\) could be dense in the Pauli basis, have large effective complexity, or be difficult to simulate. 
The main technical content of our construction is to show that one can work with a generator \(W_j\) whose norm and Pauli complexity are controlled well enough for both learning and simulation. 
We will show below that, in the logarithmically sparse regime, \(W_j\) can be supported on a Pauli space of size at most \(4^m\). 
In the polynomially sparse regime, \(W_j\) may be dense, but it admits a quasi-sparse approximation obtained from the Baker--Campbell--Hausdorff (BCH) expansion.
These structural properties are what make the formal rewriting above algorithmically useful.

It remains to explain how this generator \(W_j\) can be learned using only the allowed long-time oracle access. 
This is where the distinction between \(C_j\) and \(C_j^\dagger\) becomes important.

\noindent \textbf{Learning the correction generator $W_j$.}
We now explain how the correction generator \(W_j\) can be learned using only the allowed long-time evolution oracle access.
Recall that $W$ is chosen such that $d(C_j, e^{iW_j}) = 0$.
Here, a main difficulty is that \(C_j\) itself is not directly accessible because it requires the backward evolution \(e^{iHT}\) under the unknown Hamiltonian $H$. 
Our key observation to overcome this obstacle is that its adjoint $C_j^\dagger = e^{iH_jT}e^{-iHT}$ is accessible using only forward evolution under \(H\) and simulation of the known Hamiltonian \(H_j\). 
Since \(d(C_j,e^{iW_j})=0\), the same generator satisfies \(d(C_j^\dagger,e^{-iW_j})=0\). 
Hence, for every positive integer \(q\), we can implement, up to a global phase,
\begin{equation}
    e^{-iW_jq}
    =
    (C_j^\dagger)^q
    =
    \left(e^{iH_jT} e^{-iHT}\right)^q.
\end{equation}

Thus, long-time evolution access to \(H\) induces integer-time evolution access to the auxiliary Hamiltonian \(W_j\). 
This is precisely the type of access used in \Cref{alg:pure-Frobenius}: by querying \(e^{-iW_jq}\) for integer \(q\), we learn an approximation \(\widetilde W_j\) in normalized Frobenius norm. 

Importantly, this procedure does not violate the minimum-evolution-time constraint. 
Each application of \(C_j^\dagger\) uses one query to \(e^{-iHT}\), whose duration is exactly \(T\). 
Implementing \(e^{-iW_jq}\) simply repeats this allowed long-time query \(q\) times, so the minimum query time to the unknown dynamics remains \(t_{\min}=T\).

The efficiency of this learning step is governed by the norm and sparsity of \(W_j\). 
Moreover, the output \(\widetilde W_j\) must be produced in a sparse or quasi-sparse Pauli representation, since this is the Hamiltonian that will later be simulated to implement the correction.

\noindent \textbf{Implementing inverse time evolution of $\widetilde{W_j}$.}
For a known Hamiltonian $\widetilde{W_j}$, implementing $e^{i\widetilde{W_j}}$ is a standard task in Hamiltonian simulation. 
The circuit complexity of Hamiltonian simulation algorithms typically scales linearly with the norm and sparsity of $\widetilde{W_j}$ and polylogarithmically with the target accuracy~\cite{low2017optimal,low2019qubitization,gilyen2019qsvt,childs2012lcu,berry2015taylor,zhang2024circuit}. 
Thus, for the learned correction to be useful, it is not enough to approximate \(W_j\) accurately in norm; the estimate \(\widetilde{W_j}\) must also have bounded sparsity or quasi-sparsity.

Our learning subroutine is designed to satisfy this requirement.
It outputs \(\widetilde{W_j}\) within the sparse or quasi-sparse representation guaranteed by the structural bounds on \(W_j\). 
Consequently, the cost of simulating \(e^{i\widetilde{W_j}}\) is governed by the same norm and sparsity parameters that enter the learning of \(W_j\).

In addition, \(e^{iH_j(T+t/N)}\) can be implemented efficiently, since \(H_j\) is known, \(m\)-sparse, and satisfies the same bounded-norm assumptions as the target Hamiltonian. 
Therefore, the entire sequence in~\Cref{eq:sim-con-quantum-control} can be realized using only long-time queries to the unknown Hamiltonian \(H\), together with simulations of known Hamiltonians.

Therefore, in order to efficiently learn $W_j$ and then implement \(e^{i\widetilde W_j}\), it suffices to control the norm and sparsity, or quasi-sparsity, of the correction generator \(W_j\). 
These structural bounds ensure both that \(W_j\) can be learned from integer-time access and that the resulting estimate \(\widetilde W_j\) can be efficiently simulated. 
Concretely, we need bounds on
\begin{equation}
    \|W_j\|_F,\qquad
    \|W_j\|_\infty,\qquad
    \operatorname{supp}_P(W_j),
\end{equation}
or, in the polynomially sparse regime, on the corresponding quasi-sparsity of \(W_j\). 
These bounds are summarized in Table~\ref{tab:W-bounds} for the two operational regimes relevant to our main results. Notably, they are independent of $j$. We now turn to their derivation.
 


\begin{table}[h]
\centering
\begin{tabular}{ccccc}
\toprule
\multirow{2}{*}{$m$} 
& \multirow{2}{*}{$t_{\min}$} 
& \multicolumn{3}{c}{Properties of $W_j$} \\
\cmidrule(lr){3-5}
& & Upper bound on $\|W_j\|_F$ & Upper bound on $\|W_j\|_\infty$ & (Quasi-)sparsity \\
\midrule

$\mathcal{O}(\log n)$ 
&  $\forall\, T>0$ 
& $2\pi T\varepsilon\,\sqrt{m}$ 
& $2\pi T\varepsilon\,m$ 
& $4^m$ \\

$\mathrm{poly}(n)$ 
& $\Theta(m^{-1/K})$ 
& $\varepsilon\,\sqrt{m}$ 
& $\varepsilon\,m$ 
& $\mathcal{O}(m^{K-1})$ \\

\bottomrule
\end{tabular}
\caption{
Bounds on the correction generator $W_j$ in the two sparsity regimes considered in this work. Here, $m$ denotes the sparsity of $H$ and $H_j$, $\varepsilon$ satisfies $\|H - H_j\|_{\ell_\infty} \le \varepsilon$, and $K \in \mathbb{N}$.
}
\label{tab:W-bounds}
\end{table}

\medskip
\noindent
\textbf{Bounding the norm and (quasi-)sparsity of the auxiliary Hamiltonian $W_j$.}
We now establish the structural bounds on \(W_j\) that make the above reduction efficient. 
There are two distinct regimes of interest. 
When \(m=\mathcal{O}(\log n)\), an exponential dependence on \(m\) is still polynomial in the system size, so it suffices to show that \(W\) lies in the Pauli algebra generated by the supports of \(H\) and \(H_j\). 
This gives an exact sparsity bound of order \(4^m\). 
For polynomially sparse Hamiltonians, this exact bound is no longer efficient, and we instead prove that \(W_j\) admits a quasi-sparse approximation by truncating its BCH expansion. 
The two arguments are different, but serve the same purpose: they ensure that \(W_j\) can be learned and that the learned correction \(e^{i\widetilde W_j}\) can be simulated efficiently.


We first consider the logarithmically sparse regime, where the target Hamiltonian satisfies \(m=\mathcal{O}(\log n)\) and the minimum evolution time \(T>0\) is an arbitrary constant. 
Our goal is to bound the norm and sparsity of the auxiliary generator \(W_j\). 
In this regime, the sparsity parameter grows only logarithmically with the number of qubits, so an exponential dependence on \(m\) remains polynomial in \(n\). 
Thus, it is sufficient to prove an exact sparsity bound for \(W\), even if that bound scales exponentially in \(m\).

The key observation is that \(W_j\) can be chosen inside the finite Pauli algebra generated by the Pauli supports of \(H\) and \(H_j\). 
Indeed, the linear span of these supports over \(\mathbb F_2\) has dimension at most \(2m\), and hence contains at most \(4^m\) Pauli labels. 
This gives the exact sparsity bound \(\operatorname{supp}_P(W_j)\le 4^m\). 
The norm of \(W_j\), on the other hand, is controlled by Duhamel's principle, which relates the correction unitary \(e^{iHT}e^{-iH_jT}\) to the residual Hamiltonian \(H-H_j\).



\begin{lemma}[Informal, see Lemma~\ref{lem:sparsity-upper} for details]\label{lem:log-norm-sparsity}
Let $H,H_j$ be $n$-qubit, $m$-sparse traceless Hamiltonians satisfying $\|H-H_j\|_{\ell_\infty} \le \varepsilon$. Then there exists an $n$-qubit traceless Hamiltonian $W_j$ such that $d(e^{iHT}e^{-iH_jT}, e^{iW_j})=0$, and $\|W_j\| \le \pi T\|H-H_j\|$ and $\operatorname{supp}_P(W_j) \le 4^m$ for any unitarily invariant norm $\|\cdot\|$.
\end{lemma}

\begin{proof}[Proof Sketch]
We bound the size of the linear span generated by the Pauli supports of $H$ and $H_j$ to bound the sparsity of $W_j$. Let $C_j=e^{iHT}e^{-iH_jT}$. By Lemma~\ref{lem:log}, there exists a traceless Hermitian matrix $W_j$ satisfying $d(U,e^{iW_j})=0$ and $\|W_j\| \le \pi\|I-C_j\|$. Applying Duhamel's principle (Lemma~\ref{lem:duhamel}) yields $\|I-C_j\| \le T\|H-H_j\|$, and hence $\|W_j\| \le \pi T\|H-H_j\|$.

For the sparsity bound, let $S_H$ and $S_{H_j}$ denote the Pauli supports of $H$ and $H_j$, and define $L := \operatorname{span}_{\mathbb F_2}(S_H \cup S_{H_j}) \subseteq \mathbb F_2^{2n}$. Then $\operatorname{supp}_P(W_j) \le |L| \le 4^m$. Using $\|H-H_j\|_{\ell_\infty} \le \varepsilon$ together with the $m$-sparsity of $H$ and $H_j$, the lemma further implies that $\|W_j\|_\infty \le 2\pi mT\varepsilon$ and $\|W_j\|_F \le 2\pi \sqrt{m}\,T\varepsilon$. 
\end{proof}

We established upper bounds on the norm and sparsity of $W_j$ when the target Hamiltonian is $\mathcal{O}(\log n)$-sparse, namely
\begin{equation}
    \|W_j\|_\infty \le 2\pi m T \varepsilon, 
    \qquad
    \|W_j\|_F \le 2\pi \sqrt{m} T \varepsilon,
    \qquad
    \operatorname{supp}_P(W_j) \le 4^m.
\end{equation}
Since these quantities scale polynomially with the number of qubits $n$ (in the regime $m=\mathcal{O}(\log n)$), it follows that $W_j$ can be learned efficiently. We later present an explicit algorithm for learning $W_j$ (\Cref{alg:pure-Frobenius}).

\medskip

We next consider the regime where the Hamiltonian is $\mathrm{poly}(n)$-sparse. 
Concretely, we assume $m=\mathrm{poly}(n)$ and $T=\Theta(m^{-1/K})$ for a constant $K\in\mathbb{N}$. 
Our goal is again to bound the norm and sparsity of the correction generator $W_j$.

In this regime, the exact support bound \(\operatorname{supp}_P(W_j)\le 4^m\) from the logarithmically sparse case is no longer useful, since it is exponential in the system size. 
Thus, we cannot hope to learn \(W_j\) by treating it as an exactly sparse Pauli Hamiltonian with support \(4^m\). 
Instead, it is enough to show that \(W_j\) is quasi-sparse at the accuracy required by the learning subroutine: although \(W_j\) may be dense, it can be approximated in normalized Frobenius norm by a Hamiltonian with only polynomially many Pauli terms.

We establish this quasi-sparsity using the BCH expansion of the logarithm of \(C_j=e^{iHT}e^{-iH_jT}\). 
When \(T\) decreases as \(m^{-1/K}\), higher-order commutators are increasingly suppressed. 
Truncating the BCH series therefore yields an approximation to \(W_j\) whose Pauli support grows only polynomially with \(m\), while the truncation error remains below the accuracy needed for learning and for implementing \(e^{i\widetilde W_j}\).


\begin{lemma}[Informal, see~\Cref{prop:quasi-sparse-W} for details]\label{lem:poly-norm-sparsity}
Let $H,H_j$ be $m$-sparse traceless Hamiltonians satisfying $\|H-H_j\|_{\ell_\infty} \le \varepsilon$. Then there exists a traceless Hermitian matrix $W_j$ such that $e^{iHT}e^{-iH_jT}=e^{iW_j}$, and for any $l\in\mathbb{N}$ we have $\|W_j\| \le \|H-H_j\|$ and
\begin{equation}
\operatorname{supp}_P\!\left(W_j, \, \frac{\varepsilon m^{-\ell/K}}{8}\right) 
\le \mathcal{O}\!\left(m^{\,\ell+K/2-1}\right),   
\end{equation}
for any unitarily invariant norm $\|\cdot\|$, where $T=\Theta(m^{-1/K})$ for $K\in\mathbb{N}$.
\end{lemma}

\begin{proof}[Proof Sketch]
We express $W_j$ using the Baker--Campbell--Hausdorff (BCH) expansion (Lemma~\ref{lem:BCH}). Since $\|TH\|_\infty+\|TH_j\|_\infty \le \log 2$, the series converges.

Let $W_j^{(k)}$ denote the $k$-th order term. One can show that $\|W_j^{(k)}\|_F = \mathcal{O}(T^k \sqrt{m}\,\varepsilon)$, while its Pauli sparsity scales as $\mathcal{O}(m^k)$. Truncating the expansion at a sufficiently large order yields an approximation of $W_j$ whose effective sparsity is polynomial in $m$, establishing the claimed quasi-sparsity bound. Using $\|H-H_j\|_{\ell_\infty}\le \varepsilon$ together with the $m$-sparsity of $H$ and $H_j$, the lemma further implies that $\|W_j\|_\infty \le m\varepsilon$ and $\|W_j\|_F \le \sqrt{m}\,\varepsilon$.
\end{proof}

We therefore obtain the bounds
\begin{equation}
    \|W_j\|_\infty \le m \varepsilon, 
    \qquad
    \|W_j\|_F \le \sqrt{m} \varepsilon,
    \qquad
    \operatorname{supp}_P \left(W_j, \,\frac{\varepsilon}{8\sqrt m} \right) \le \mathcal{O}(m^{K-1}).
\end{equation}
Thus, although \(W_j\) need not be exactly sparse, it can be replaced at the required precision by a polynomially sparse Hamiltonian. 
This is sufficient both for learning \(W_j\) using \Cref{alg:pure-Frobenius} and for simulating the learned correction \(e^{i\widetilde W_j}\). 
The price is the tradeoff \(T=\Theta(m^{-1/K}), K\in \mathbb{N}\),  which relates the minimum accessible evolution time to the polynomial overhead in total evolution time. For every constant $K=\mathcal{O}(1)$, the quasi-sparsity of $W_j$ becomes a polynomial with $m$.


Since these quantities (norm, sparsity) scale polynomially with the number of qubits $n$ in both regimes, it follows that $W_j$ can be learned efficiently. We now analyze the complexity of the algorithm for learning $W_j$ (\Cref{alg:pure-Frobenius}).

\medskip
\noindent
\textbf{Complexity of learning the auxiliary Hamiltonian $W_j$.}
We have shown that the norm and sparsity of $W_j$ admit polynomial bounds. 
Under these guarantees, standard Hamiltonian learning algorithms require only polynomial total evolution time. 
Our Hamiltonian learning algorithm (\Cref{alg:pure-Frobenius}) for the auxiliary Hamiltonian $W_j$ likewise achieves polynomial complexity. 
In this section, we explicitly compute the total evolution time required to estimate $W_j$.

To this end, we first specify the constant $c$, which controls the target accuracy. 
For simplicity, we set $c=\mathcal{O}(1/\sqrt{m})$ (see Appendix~\ref{sec:app-main} for details). 
The remaining parameters $c_F$, $c_\infty$, and $s$ in \Cref{alg:pure-Frobenius} correspond to upper bounds on $\|W_j\|_F$, $\|W_j\|_\infty$, and the (quasi-)sparsity of $W_j$, respectively, as derived in the previous section and summarized in Table~\ref{tab:W-bounds}.

The parameters in Algorithm~\ref{alg:pure-Frobenius} are given by
\begin{equation}
    t = \left\lfloor \frac{c}{10c_Fc_\infty\varepsilon} \right\rfloor,
    \qquad 
    \delta_t = c_F c_\infty t^2 \varepsilon^2.
\end{equation}
Since $t$ must be a positive integer, we require \(\varepsilon \le {c}/{(10c_Fc_\infty)} = \eta_{\mathrm{sw}}.\) For the complementary regime $\varepsilon \ge \eta_{\mathrm{sw}}$, we present an alternative algorithm in Appendix~\ref{sec:sql}.

The algorithm requires 
\(M = \mathcal{O}\!\left(s/\delta_t^2\right)\)
queries to $e^{-iW_jt}$. 
Each query to $e^{-iW_jt}$ corresponds to evolution time $\mathcal{O}(tT)$ of $H$, so the total evolution time is $MtT=\mathcal{O}(stT/\delta_t^2)$. 
Substituting the values of the parameters $c,c_F,c_\infty,s$, we obtain the following bounds. For logarithmically sparse Hamiltonians,
\begin{equation}\label{eq:log-control}
    MtT = \widetilde{\mathcal{O}}\!\left(\frac{4^m T^3}{\varepsilon}\right).
\end{equation}
In particular, this shows that the minimum accessible evolution time can be chosen as an arbitrary constant. For polynomially sparse Hamiltonians, we obtain
\begin{equation}\label{eq:poly-control}
    MtT = \widetilde{\mathcal{O}}\!\left(\frac{m^{K+2} T}{\varepsilon}\right).
\end{equation}
This demonstrates that the minimum evolution time $T$ can be relaxed from $\Theta(m^{-1})$ to $\Theta(m^{-1/K})$ at the cost of an increased total evolution time. Overall, our approach reveals a quantitative tradeoff between the minimum evolution time and the total evolution time required to approximate continuous quantum control in Hamiltonian learning.

Our framework learns the unitary $C_j$ in order to approximate continuous quantum control. In principle, this task could also be approached via general unitary channel estimation~\cite{haah2023query}. However, producing an $\varepsilon$-accurate estimate of an arbitrary $n$-qubit unitary requires $\mathcal{O}(4^n/\varepsilon)$ queries, which is exponential in the system size even when the underlying Hamiltonian is sparse. In contrast, our framework exploits the specific structure that $C_j$ is generated by the time evolution of a logarithmically sparse Hamiltonian, and therefore achieves the same task using only a polynomial number of queries. This highlights a key advantage of our approach and underscores the importance of leveraging dynamical structure in Hamiltonian learning.

The discussion above constructs the residual-evolution primitive required at each refinement step. 
Once this primitive is available, the remaining task is to convert residual dynamics into coefficient estimates. 
This is achieved by a separate reduction to sparse pure-state tomography, which we describe next.

\begin{algorithm}[t]
\caption{$\textsc{IntegerEvol}(\mathcal{U}_A,s,c_F,c_\infty,c,\varepsilon)$: Hamiltonian learning in normalized Frobenius norm via pure-state tomography using integer time evolution}
\label{alg:pure-Frobenius}
\DontPrintSemicolon

\KwIn{Query access to $\mathcal{U}_A(\tau)=e^{-iA\tau}, \tau\in\mathbb{N}$ for a traceless Hermitian operator $A=\sum_x \alpha_x P_x$ satisfying $\|A\|_F\le c_F\varepsilon$ and $\|A\|_\infty\le c_\infty\varepsilon$; target constant $c$; effective support size $s=\operatorname{supp}_P(A,c\varepsilon/8)$.}

\KwOut{An estimate $\widetilde A$ such that $\|A-\widetilde A\|_F\le c\varepsilon$.}

$t \gets \left\lfloor {c}/{(10c_Fc_\infty\varepsilon)}\right\rfloor$\;

$\delta_t \gets c_Fc_\infty t^2\varepsilon^2$\;

Prepare the maximally entangled state $\ket{\Phi}$\;

Let $V$ be a unitary such that $V(P_x\otimes I)\ket{\Phi}=\ket{x}$ for all $x$\;

Prepare $\ket{\psi_t}\gets V(\mathcal{U}_A(t)\otimes I)\ket{\Phi}$\;

\tcp{From the first-order expansion of $e^{-iAt}$, $\left\|\ket{\psi_t}-\ket{0}+it\sum_x \alpha_x\ket{x}\right\|_2
\le \delta_t$.}

Apply sparse pure-state tomography (Lemma~\ref{lem:sparse-tomo-l2}) to learn $\ket{\psi_t}$ with accuracy $\delta_t$ using $\mathcal{O}({s}/{\delta_t^{2}})$ copies, and obtain coefficients $\{\widetilde{\alpha}_x\}_x$ satisfying
\[
\left(\sum_x \left|\alpha_x-\widetilde{\alpha}_x \right|^2\right)^{1/2}\le c\varepsilon .
\]

\Return{$\widetilde A \gets \sum_x \widetilde{\alpha}_x P_x$}\;

\end{algorithm}

\begin{algorithm}[t]
\caption{$\textsc{SparseHam}(\mathcal{U}_A,m,\varepsilon)$: Sparse Hamiltonian learning in $\ell_\infty$ norm via pure-state tomography}
\label{alg:pure-l-infty}
\DontPrintSemicolon

\KwIn{Query access to $\mathcal{U}_A(\tau)=e^{-iA\tau}$ for a traceless Hamiltonian $A=\sum_x \alpha_x P_x$ satisfying $\|A\|_{\ell_\infty}\le \varepsilon$ and $m=\operatorname{supp}_P(A)$.}

\KwOut{An estimate $\widetilde A$ such that $\|A-\widetilde A\|_{\ell_\infty}\le \varepsilon/8$.}

$t \gets 1/ (32m\varepsilon)$\;

Prepare the maximally entangled state $\ket{\Phi}$\;

Let $V$ be a unitary such that $V(P_x\otimes I)\ket{\Phi}=\ket{x}$ for all $x$\;

Prepare $\ket{\psi_t} \gets V(\mathcal{U}_A(t)\otimes I)\ket{\Phi}$\;

\tcp{From the first-order expansion of $e^{-iAt}$, $\left\|\ket{\psi_t}-\ket{0}+it\sum_x \alpha_x\ket{x}\right\|_{\ell_\infty}
\le mt^2\varepsilon^2$.}

Apply sparse pure-state tomography (Lemma~\ref{lem:sparse-tomo-linf}) to learn $\ket{\psi_t}$ with accuracy $mt^2\varepsilon^2$ using $\mathcal{O}(m^3)$ copies, and obtain coefficients $\{\widetilde{\alpha}_x\}_x$ satisfying $|\alpha_x-\widetilde{\alpha}_x|\le \varepsilon/8$ for all $x$

\Return{$\widetilde A \gets \sum_x \widetilde{\alpha}_x P_x$}\;
\end{algorithm}

\section{Coefficient extraction by pure-state tomography}
\label{sec:tomo-extraction}

We now explain how the emulated residual dynamics is converted into coefficient estimates. 
The basic idea is an amplitude encoding of Pauli coefficients into a sparse quantum state through the first-order expansion of the time evolution operator. 
Recall that there are two uses of pure-state tomography in our protocol. 
The auxiliary use, described above, learns the correction generator \(W_j\) needed to emulate residual dynamics. 
The primary use, described in this section, extracts the coefficients of the residual Hamiltonian \(\Delta H_j=H-H_j\) once this emulated dynamics is available. 
Both uses rely on the same first-order amplitude-encoding idea: by applying an evolution \(e^{-iAt}\) to a maximally entangled state, the Pauli coefficients of \(A\) are encoded in the amplitudes of the resulting quantum state.

We now describe this reduction in a unified form. 
Let $A=\sum_x \alpha_x P_x$ be a Hamiltonian with $P_0=I$ and $\alpha_0=0$, and define the Pauli sparsity $m=\operatorname{supp}_P(A)$. Suppose that $\|A\|_{\ell_\infty}\le\varepsilon$ and that we can implement the time-evolution operator $e^{-iAt}$. For sufficiently small $t$, we have the expansion $e^{-iAt}=I-iAt+\mathcal{O}(t^2\varepsilon^2)$.

Using this evolution, we prepare the state $\ket{\psi_t}=V(e^{-iAt}\otimes I)\ket{\Phi}$, where $\ket{\Phi}$ is a maximally entangled state and $V$ is a unitary satisfying $(P_x\otimes I)\ket{\Phi}\mapsto \ket{x}$. A convenient choice is the modified Bell-basis decoding circuit $V := \bigotimes_{j=1}^n\left(H_{S_j}\otimes I_{A_j}\right)\left(\operatorname{CNOT}_{S_j\to A_j}\right),$    
for which, writing $x=(a,b)\in\{0,1\}^{2n}$ and $P_x = i^{a\cdot b}X^a Z^b$, we have $V(P_x\otimes I)\ket{\Phi}_{SA} = \ket{b}_S \ket{a}_A \equiv \ket{x}$.

Substituting the expansion of $e^{-iAt}$ yields $\ket{\psi_t}\simeq \ket{0}-it\sum_x \alpha_x\ket{x}$. Hence, the Pauli coefficients $\alpha_x$ can be recovered via pure-state tomography of $\ket{\psi_t}$. More precisely, the approximation error satisfies
\begin{equation}
    \left\|\ket{\psi_t}-\ket{0}+it\sum_x\alpha_x\ket{x}\right\|_{\ell_\infty}
\le mt^2\|A\|_{\ell_\infty}^2 \le mt^2\varepsilon^2.
\end{equation}

Learning $\ket{\psi_t}$ to accuracy $\mathcal{O}(mt^2\varepsilon^2)$ yields an estimate of $t\alpha_x$ with the same accuracy, and hence an estimate of $\alpha_x$ with error $\mathcal{O}(mt\varepsilon^2)$. Choosing $t=\Theta(1/(m\varepsilon))$ ensures this error is $\mathcal{O}(\varepsilon)$. Therefore, it suffices to learn $\ket{\psi_t}$ to accuracy $\mathcal{O}(1/m)$.

Since $A$ is (quasi-)sparse, the state $\ket{\psi_t}$ is also (quasi-)sparse with sparsity parameter $m$. Sparse pure-state tomography with accuracy $\mathcal{O}(1/m)$ requires $S=\mathcal{O}(m^3)$ samples, yielding a total evolution time of $\mathcal{O}(St)=\mathcal{O}(m^2/\varepsilon)$, which matches the Heisenberg limit. A detailed analysis, including extensions to other norms such as the normalized Frobenius norm, is provided in Appendix~\ref{sec:repure}.
We now combine this coefficient-extraction primitive with the long-time emulation primitive to obtain the full iterative learner and its complexity.

\section{Full learning protocol and complexity}

\label{sec:full-protocol}

\begin{algorithm}[t]
\caption{\textsc{Main}: Hamiltonian learning in Heisenberg limit without short time control}
\label{alg:main}
\DontPrintSemicolon

\SetKwFunction{LearnF}{IntegerEvol}
\SetKwFunction{LearnSQL}{LearnStandardQuantumLimit}
\SetKwFunction{LearnS}{SparseHam}
\SetKwFunction{Thresh}{T_m}

\KwIn{Oracle access to $e^{-iH\tau}$ for all $\tau\ge T$, where $H$ is traceless, $m$-sparse, and $\|H\|_{\infty} \le 1$; target accuracy $\varepsilon\in(0,1)$.}
\KwOut{An $m$-sparse estimate $\widetilde H$ satisfying $\|H-\widetilde H\|_{\ell_\infty}\le \varepsilon$.}

Choose $J=\lceil \log_2(1/\varepsilon)\rceil$\;

\If{$m=\mathcal{O}(\log n)$}{
    $s \gets 4^m$, \quad
    $c_F \gets 2\pi \sqrt{m}T$, \quad
    $c_\infty \gets 2\pi mT$, \quad
    $c \gets {1}/{(256\sqrt{m})}$
}
\ElseIf{$m=\mathrm{poly}(n)$ and $T=\Theta(m^{-1/K})$ for some $K\in\mathbb{N}$}{
    $s \gets \mathcal{O}(m^{K-1})$, \quad
    $c_F \gets 2\sqrt{m}$, \quad
    $c_\infty \gets 2m$, \quad
    $c \gets {1}/{(256\sqrt{m})}$
}

$\eta_{\mathrm{sw}} \gets c/(10c_Fc_\infty)$\;

Initialize $H_0\gets 0$\;

\For{$j=0$ \KwTo $J-1$}{
    $\eta_j \gets 2^{-j}$,\quad
    $t_j \gets {1}/{(32m\eta_j)}$,\quad
    $N_j\gets \lceil{1}/{(2\sqrt{m}\eta_j)}\rceil$\;

    $C_j \gets e^{iHT} e^{-iH_jT}$\;

    \eIf{$\eta_j \le \eta_{\mathrm{sw}}$}{
        Let $W_j$ be Hermitian such that $C_j=e^{iW_j}$\;

        Implement integer-time access to $\mathcal{U}_{W_j}(t)=e^{-iW_jt}$ via $(C_j^{\dagger})^t$\;

        $\widetilde W_j \gets \textsc{IntegerEvol}(\mathcal{U}_{W_j},s,c_F,c_\infty,c,\eta_j)$\;

        $\widetilde C_j \gets e^{i\widetilde W_j}$\tcp*{$\|W_j-\widetilde W_j\|_F\le c\eta_j$}

        $\tau_j \gets T+t_j/N_j$\;

        $\widetilde U_{\mathrm{res}}^{(j)} 
        \gets \left(e^{-iH\tau_j}\,\widetilde C_j\,e^{iH_j\tau_j}\right)^{N_j}$\tcp*{$\approx e^{-i(H-H_j)t_j}$}

        $\widetilde{\Delta H}_j \gets 
        \textsc{SparseHam}\left(\widetilde U_{\mathrm{res}}^{(j)},m,\eta_j\right)$\tcp*{$\|\widetilde{\Delta H}_j-\Delta H_j\|_{\ell_\infty}\le \eta_j/4$}
    }{
        $\widetilde{\Delta H}_j \gets 
        \textsc{LearnStandardQuantumLimit}(\mathcal{U}_H,H_j,m,\eta_j)$\tcp*{$\|\widetilde{\Delta H}_j-\Delta H_j\|_{\ell_\infty}\le \eta_j/4$}
    }

    $H_{j+1} \gets \mathcal{T}_{m,1} \left(H_j+\widetilde{\Delta H}_j\right)$\tcp*{ensures $\|H_{j+1}\|_{\infty} \le 1$ and $\|H-H_{j+1}\|_{\ell_\infty}\le \eta_j/2$}
}

\Return{$\widetilde H \gets H_J$}\;

\end{algorithm}
We now assemble the components from the previous sections into the full Hamiltonian learning protocol and derive the complexity bounds in \Cref{thm:arb-long,thm:many-long}.
Recall from the algorithmic overview that the protocol follows an iterative refinement strategy. 
At the beginning of round \(j\), we have an estimate \(H_j\) satisfying
\begin{equation}
    \|H-H_j\|_{\ell_\infty}\le \eta_j .
\end{equation}
The round learns the residual Hamiltonian
\begin{equation}
    \Delta H_j := H-H_j
\end{equation}
to accuracy proportional to \(\eta_j\), and then updates \(H_j\) so that the error is reduced by a constant factor.

In the small-error regime, one refinement round has the concrete form
\begin{equation}
H_j
\;\xrightarrow{\textsc{IntegerEvol}}
\widetilde W_j
\;\xrightarrow{\text{long-time product formula}}
\widetilde U_{\mathrm{res}}^{(j)}
\;\xrightarrow{\textsc{SparseHam}}
\widetilde{\Delta H}_j
\;\xrightarrow{\mathcal T_{m,1}}
H_{j+1}.
\end{equation}
Here \textsc{IntegerEvol} learns the correction generator needed for the long-time emulation primitive developed in \Cref{sec:long-time-emulation}. 
The resulting long-time product formula implements the residual dynamics, which is then converted into coefficient estimates by the pure-state tomography reduction described above. 
The final truncation step restores the promise that the estimate remains \(m\)-sparse and norm bounded.

We now explain the main ingredients of \Cref{alg:main}. 
The algorithm maintains an estimate \(H_j\) inside the feasible set
\begin{equation}
    \mathcal{S}_{k,c}:=\{H_0:\operatorname{supp}_P(H_0)\le k,\ \|H_0\|_\infty\le c\}.
\end{equation}
To enforce this condition after each update, we use the sparse bounded truncation map
\begin{equation}
    \mathcal{T}_{k,c}(H):=
    \operatorname*{arg\,min}_{H_0\in\mathcal{S}_{k,c}}
    \|H-H_0\|_{\ell_\infty}.
\end{equation}
Starting from \(H_0=0\), the algorithm aims to maintain the invariant
\begin{equation}
    \|H-H_j\|_{\ell_\infty}\le \eta_j,\qquad \eta_j=2^{-j}.
\end{equation}
Thus, each round only needs to learn the residual \(\Delta H_j=H-H_j\) well enough to halve the current error.
After \(J=\lceil \log_2(1/\varepsilon)\rceil\) rounds, the estimate \(H_J\) reaches the target accuracy.

The algorithm has two regimes. 
When \(\eta_j\) is small enough, we use the Heisenberg-limit subroutine based on long-time control emulation. 
For earlier rounds, when \(\eta_j\) is still relatively large, we use a simpler subroutine. 
The two regimes are separated by the crossover point \(\eta_{\mathrm{sw}}\) defined in \Cref{alg:main}.

\medskip
\noindent
\textbf{Small error regime ($\eta_j \le \eta_{\mathrm{sw}}$).}  
When the current residual error is sufficiently small, we operate in a regime based on long-time control emulation. 
This regime is responsible for achieving the final Heisenberg-limited scaling in the target accuracy. 
Intuitively, once we are close enough to the true Hamiltonian, long-time evolution suffices to emulate residual dynamics and extract higher-precision information about the Hamiltonian.  
Each round consists of the following three steps.

\begin{enumerate}[(1)]
\item \textbf{Quantum control emulation (\Cref{alg:pure-Frobenius}).}  
The goal of this step is to coherently simulate the residual evolution $e^{-i\Delta H_j t_j}$ without access to short-time queries. 
To this end, we first learn an effective Hamiltonian $W_j$ that captures the discrepancy between the true dynamics and the current hypothesis. 
We then construct an approximate unitary $\widetilde{U}_{\mathrm{res}}^{(j)}$ via a product formula using only long-time evolutions:
\begin{equation}
    \widetilde{U}_{\mathrm{res}}^{(j)} = \left(e^{-iH(T+t_j/N_j)}e^{i\widetilde{W}_j}e^{iH_j(T+t_j/N_j)}\right)^{N_j}.    
\end{equation}
Here $\widetilde{W}_j$ approximates $W_j$, which is defined implicitly through the exact identity 
\begin{equation}
    d(e^{-iW_j},e^{iH_jT}e^{-iHT})=0.
\end{equation}
This construction can be interpreted as a stroboscopic simulation that cancels the known component $H_j$ while isolating and amplifying the residual $\Delta H_j$.

Using \Cref{alg:pure-Frobenius} with $c=\mathcal{O}(1/\sqrt{m})$, we obtain
\[
\|W_j-\widetilde{W}_j\|_F = \mathcal{O}(\eta_j/\sqrt{m}).
\]
This error accumulates linearly in the number of repetitions $N_j$, yielding
\begin{equation}
\left\|\widetilde{U}_{\mathrm{res}}^{(j)}-e^{-i\Delta H_j t_j}\right\|_F = \mathcal{O}\left(\frac{N_j\eta_j}{\sqrt{m}}\right) = \mathcal{O} \left(\frac1m\right).
\end{equation}
Thus, by choosing $N_j$ appropriately, we ensure inverse-polynomial accuracy in the implemented residual evolution.

\item \textbf{Coefficient and structure learning (\Cref{alg:pure-l-infty}).}  
Having obtained an accurate implementation of the residual evolution, we convert this dynamical information into a state encoding the coefficients of $\Delta H_j$. 
Specifically, we prepare
\begin{equation}
    \ket{\psi_j} = V\left(e^{-i\Delta H_j t_j}\otimes I\right)\ket{\Phi} \simeq \ket{0} - it_j \sum_{x\neq 0}\alpha_x^{(j)}\ket{x},    
\end{equation}
where $t_j = \mathcal{O}(1/(m\eta_j))$.
This approximation follows from the first-order expansion of the evolution, valid since $t_j$ is sufficiently small relative to $\|\Delta H_j\|_{\ell_\infty}$.

Applying sparse tomography, we estimate these coefficients. 
Writing $\ket{\psi_j}=\sum_x \beta_x\ket{x}$, the guarantee
\[
|\beta_x + it_j \alpha_x^{(j)}| = \mathcal{O}(1/m)
\]
implies that each $\alpha_x^{(j)}$ can be estimated within error $\mathcal{O}(1/(mt_j)) = \eta_j/4$. 
This yields an estimate $\widetilde{\Delta H}_j$ satisfying
\[
\|H-(H_j+\widetilde{\Delta H}_j)\|_{\ell_\infty} \le \eta_j/4.
\]
Importantly, this step recovers both the support and the coefficients of the residual Hamiltonian.

\item \textbf{Sparse bounded truncation (\Cref{lem:sbt-stab}).}  
We update the hypothesis via
\[
H_{j+1} = \mathcal{T}_{m,1}(H_j + \widetilde{\Delta H}_j).
\]
The truncation operator $\mathcal{T}_{m,1}$ enforces that the estimate remains $m$-sparse and has norm at most $1$, thereby preventing error accumulation and inefficiency across iterations.
By stability of sparse bounded truncation, we obtain
\[
\|H-H_{j+1}\|_{\ell_\infty} \le \eta_j/2.
\]
Thus, each refinement round reduces the error by a constant factor while preserving the structural constraints.
\end{enumerate}

The total evolution time in the small error regime is 
$t_{\mathrm{tot,small}}^{(j)} = \widetilde{\mathcal{O}}(4^m T^3 / \eta_j)$ for logarithmically sparse Hamiltonians and 
$\widetilde{\mathcal{O}}(m^{K+2} T / \eta_j)$ for polynomially sparse Hamiltonians. 
In the former case, $T$ can be chosen as an arbitrarily large constant. 
In the latter case, we require $T = \Theta(m^{-1/K})$ for some constant $K\in\mathbb{N}$ to ensure that the simulation remains well-conditioned. 
See Appendix~\ref{sec:app-main} for detailed derivations.

\medskip
\noindent
\textbf{Large error regime ($\eta_j \ge \eta_{\mathrm{sw}}$).}
The long-time control-emulation step above is used only once the residual error is sufficiently small. 
For larger \(\eta_j\), the integer-time learning parameter in \Cref{alg:pure-Frobenius} would be too small to yield a useful Heisenberg-limited primitive. 
In these early rounds, we instead use the simpler subroutine \Cref{alg:sql}.

In this regime, we avoid quantum control emulation and instead directly learn $\mathcal{U}_{A,A_0}(t)$ using \Cref{alg:sql}. 
The unitary is reconstructed via sparse pure-state tomography, from which we extract $\widetilde{\Delta H}_j$ using its first-order expansion. 

Compared to the small error regime, this approach incurs a worse dependence on $\eta_j$ but avoids the overhead of simulating continuous control. 
After truncation, we again obtain $H_{j+1} \in S_{m,1}$ with $\|H-H_{j+1}\|_{\ell_\infty} \le \eta_j/2$. 

The total evolution time in the larger error regime is $t_{\mathrm{tot,large}}^{(j)} = \widetilde{\mathcal{O}}(4^m T / \eta_j^2)$ for logarithmically sparse Hamiltonians and 
$\widetilde{\mathcal{O}}(m^K T / \eta_j^2)$ for polynomially sparse Hamiltonians. 
In the former case, $T$ can be chosen as an arbitrarily large constant, whereas in the latter case we require $T = \Theta(m^{-1/K})$ for some constant $K\in\mathbb{N}$. See Appendix~\ref{sec:app-main} for detailed derivations.

\medskip
\noindent
\textbf{Total complexity and regime crossover.}
At the crossover point $\eta_j = \eta_{\mathrm{sw}}$ (defined in \Cref{alg:main}), the small error regime and large error regime complexities coincide, and each is optimal in its respective domain. Hence $t_{\mathrm{tot}}^{(j)} = \min\{t_{\mathrm{tot,small}}^{(j)}, t_{\mathrm{tot,large}}^{(j)}\}$ and $t_{\mathrm{tot}} = \sum_{j=1}^J t_{\mathrm{tot}}^{(j)}$.

For logarithmically sparse Hamiltonians, we obtain
\begin{equation}
    t_{\mathrm{tot}} = \widetilde{\mathcal{O}}\!\left(
\min\left\{\frac{4^m T^3}{\varepsilon}, \frac{4^m T}{\varepsilon^2}\right\}
\right),    
\end{equation}
with minimum evolution time $t_{\min} = T$.

For polynomially sparse Hamiltonians, we obtain
\begin{equation}
    t_{\mathrm{tot}} = \widetilde{\mathcal{O}}\!\left(
\min\left\{\frac{m^{K+2}T}{\varepsilon}, \frac{m^K T}{\varepsilon^2}\right\}
\right),    
\end{equation}
with $t_{\min} = \Theta(m^{-1/K}) = T$ for $K\in\mathbb{N}$. This establishes \Cref{thm:arb-long,thm:many-long}.

The resulting bounds show that the minimum accessible evolution time is not merely a technical parameter of the algorithm, but a resource that can be traded against total evolution time. We now discuss the implications of this tradeoff and several directions opened by the long-time learning model.

\section{Discussion}
To the best of our knowledge, existing Heisenberg-limited protocols for learning an \(m\)-sparse Hamiltonian without continuous quantum control rely on access to very short evolution times, with the minimum evolution time scaling as \(t_{\min}=\Theta(\sqrt{\varepsilon})\)~\cite{abbas2025nearly} or \(t_{\min}=\Theta(1/m)\)~\cite{bakshi2024learning}. 
In sharp contrast, our framework demonstrates that Heisenberg-limited Hamiltonian learning is possible without access to very short-time evolutions, thereby eliminating the need for ultra-short-time control in the logarithmically sparse regime and substantially relaxing it in the polynomially sparse regime.
This represents a qualitative shift in the paradigm of Hamiltonian learning: rather than relying on fine-grained temporal access, one can instead leverage long-time dynamics together with suitable algorithmic reductions to achieve information-theoretically optimal performance. 
In particular, for logarithmically sparse Hamiltonians, our results show that Heisenberg-limited scaling can still be attained even when the minimum accessible evolution time is fixed to be a constant independent of the system size and target accuracy.

We highlight the fundamental open problem of characterizing the optimal tradeoff between the total evolution time $t_{\mathrm{tot}}$ and the minimum evolution time $t_{\min}$. 
Can one design algorithms that achieve $t_{\mathrm{tot}} = \mathrm{poly}(m,T)/\varepsilon$ while maintaining $t_{\min}=T$? 
More broadly, what is the optimal dependence of $t_{\mathrm{tot}}$ on the sparsity $m$ and the minimum evolution time $T$? 
Resolving these questions would provide a complete understanding of the role of temporal resources in Hamiltonian learning.

This conceptual shift brings crucial clarity to the distinction between minimum evolution time and time resolution.
While the minimum evolution time ($t_{\min} := \min_{i} t_i$) is often referred to as the minimum time resolution in the literature~\cite{bakshi2024learning, abbas2025nearly, huang2023learning}, from a hardware perspective, it is more natural to formalize time resolution as a fundamental discretization step $\delta$ (e.g., the clock speed of a control generator).
Under this definition, accessible evolutions are restricted to $U(t_j) = e^{-iHt_j}$ with $t_j \in \delta \mathbb{N}$, even if all such times are required to satisfy $t_j\ge t_{\min}$. 
Thus, $t_{\min}$ and $\delta$ capture distinct experimental limitations: the former specifies the shortest duration for which the system must be coherently evolved, whereas the latter specifies how finely the evolution time must be programmable. 
Our results address the former by showing that Heisenberg-limited learning can be achieved without accessing ultra-short evolutions. 
Determining the minimum time resolution $\delta$ required for efficient Hamiltonian learning remains an important open problem.


Beyond temporal constraints, a crucial practical direction is to study Hamiltonian learning under the limited availability of ancillary systems.
In many physical platforms, introducing and maintaining clean ancillas is highly restricted or costly.
Inspired by recent progress in ancilla-free Hamiltonian learning~\cite{hu2025ansatz}, it would be highly desirable to develop ancilla-free algorithms that operate solely using long-time evolutions. 
Furthermore, understanding the interplay between long-time dynamics and additional control resources, such as static single-qubit fields~\cite{brahmachari2026learning}, may provide new avenues for improving practical performance while maintaining theoretical guarantees.

From an experimental perspective, several additional directions merit investigation. One natural question concerns even more restrictive access models, where only a limited subset of the system can be probed, without requiring global control over the dynamics. In recent work on quantum probe tomography~\cite{chen2025probe}, it has been shown that, for certain structured Hamiltonians, Heisenberg-limited scaling can be achieved even under such locality constraints, using only local probe operations. It is therefore natural to ask whether similar guarantees can be established in the long-time evolution regime considered in this work. In particular, can one achieve Heisenberg-limited Hamiltonian learning using only local probes combined with long-time dynamics, thereby completely eliminating the need for global control?

Another important experimental consideration concerns the precision with which long-time evolution can be implemented. While our framework removes the need for ultra-short-time control, it still implicitly assumes the ability to realize time evolution for a prescribed duration $T$ with high accuracy. In realistic experimental settings, however, such timing may only be achievable up to multiplicative uncertainties, e.g., implementing an evolution time in the range $[(1-\delta)T, (1+\delta)T]$ for some constant $\delta > 0$. It is therefore an important open question whether Heisenberg-limited scaling can still be achieved under such timing imprecision. Addressing this question would significantly enhance the robustness and practical relevance of long-time Hamiltonian learning protocols.

\section*{Acknowledgments}
M.S. thanks Yu Tong for insightful discussions. J.L. thanks Anurag Anshu, Sitan Chen, Francisco Escudero Gutiérrez, Weiyuan Gong, Antonio Anna Mele, and Chirag Wadhwa for helpful feedback on an earlier draft and for insightful discussions. J.L. also thanks Constantin Cedillo Vayson de Pradenne, Jordan Cotler, and Hsin-Yuan Huang for sharing their concurrent work.

J.L. acknowledges support from the National Research Foundation of Korea (NRF) grants funded by the MSIT (Grant No. RS-2025-00515537) and the Institute for Information \& Communications Technology Promotion (IITP) grant funded by the Korean government (MSIP) (Grant Nos. RS-2025-02304540 and RS-2019-II190003). C.O. was supported by the NRF Grants (No. RS-2024-00431768 and No. RS-2025-00515456) funded by the Korean government (MSIT) and IITP grants funded by the Korea government (MSIT) (No. RS-2024-00437284, No. IITP-2025-RS-2025-02283189 and No. IITP-2025-RS-2025-02263264) and by Global Partnership Program of Leading Universities in Quantum Science and Technology (RS-2025-08542968) through the National Research Foundation of Korea~(NRF) funded by the Korean government (Ministry of Science and ICT(MSIT)).




\bibliographystyle{unsrt}
\let\oldaddcontentsline\addcontentsline
\renewcommand{\addcontentsline}[3]{}
\bibliography{citation}
\let\addcontentsline\oldaddcontentsline

\newpage
\appendix
\appendixpage

\tableofcontents

\section{Related work}
The interactions of a quantum system are encoded in its Hamiltonian, and characterizing this Hamiltonian from real-time dynamics is a fundamental problem in quantum information and many-body physics. In particular, learning an unknown Hamiltonian from access to its time-evolution operator has been extensively studied; see, e.g.,~\cite{huang2023learning, yu2023robust, ma2024learning, hu2025ansatz, abbas2025nearly, gu2024practical, li2024heisenberg, zhao2025learning, da2011practical, wiebe2014hamiltonian, dutkiewicz2024advantage, wang2017experimental, pang2014quantum, shaffer2021practical, zubida2021optimal}. Beyond full reconstruction, more recent works have also considered weaker primitives such as Hamiltonian testing~\cite{aharonov2022quantum, laborde2022quantum, she2022unitary, bluhm2024hamiltonian, arunachalam2025testing, kallaugher2025hamiltonian, gao2025quantum} and certification~\cite{gao2025quantum, bluhm2025certifying, lee2025optimal, bluhm2026certifying}, which aim to verify structural or spectral properties of a Hamiltonian without fully learning it.

In this work, we focus on learning the full description of an unknown Hamiltonian using only real-time dynamics. It is by now well understood that Hamiltonian learning can be performed quite generally using derivative-based estimation techniques, which achieve total evolution time scaling $t_{\mathrm{tot}} = \mathcal{O}(\log(n)/\varepsilon^3)$~\cite{zubida2021optimal}. 

More efficient guarantees are known under additional structural assumptions. In particular, when the interaction structure is known and the Hamiltonian is low-interaction (e.g., bounded-degree local), one may assume that the Pauli operators $P_x$ are fixed and only the coefficients $\alpha_x$ are unknown. In this setting, Huang, Tong, Fang, and Su~\cite{huang2023learning} show that the Hamiltonian can be learned to accuracy $\varepsilon$ using total evolution time $t_{\mathrm{tot}} = \Theta(\log(n)/\varepsilon)$ with minimum evolution time $t_{\min} = \Theta(\sqrt{\varepsilon})$, achieving Heisenberg-limited scaling in the error parameter and improving upon the standard $1/\varepsilon^2$ dependence.

However, the assumption that the interaction structure is known is often unrealistic in practice. Existing approaches relying on this assumption typically require explicit knowledge of the interaction terms, either to construct commutator expansions~\cite{haah2022optimal} or to perform Hamiltonian reshaping procedures that decouple the dynamics~\cite{huang2023learning}. 

Learning the interaction structure itself has therefore attracted significant attention~\cite{bakshi2024structure, abbas2025nearly, hu2025ansatz}. In particular, for $\mathcal{O}(1)$-local Hamiltonians, Bakshi, Liu, Moitra, and Tang~\cite{bakshi2024structure} construct Goldreich--Levin-type queries using classical shadows, enabling efficient identification of the support. Their approach achieves Heisenberg-limited total evolution time $t_{\mathrm{tot}} = \mathcal{O}(m/\varepsilon)$ via a hierarchical learning scheme, while maintaining a $\varepsilon$-independent minimum evolution time $t_{\min} = \Theta(1/m)$ for general local Hamiltonians and $t_{\min} = \Theta(1)$ for local Hamiltonians when every site is acted on by $\mathcal{O}(1)$ terms, through approximate quantum control implemented using constant-time Trotterization. Since the fundamental lower bound is $t_{\mathrm{tot}} = \Omega(1/\varepsilon)$~\cite{dutkiewicz2024advantage}, this result is optimal in its dependence on $\varepsilon$ for local Hamiltonians.

Hamiltonian learning without locality assumptions has also been explored~\cite{hu2025ansatz, abbas2025nearly}. The ansatz-free approach of~\cite{hu2025ansatz} achieves total evolution time $t_{\mathrm{tot}} = \mathcal{O}(m^2/\varepsilon)$ via a hierarchical learning scheme combined with Hamiltonian reshaping. This method isolates individual Pauli components using Pauli twirling, allowing efficient estimation of coefficients. However, implementing the reshaped dynamics requires repeated applications of ultra-short time evolutions, resulting in a minimum evolution time of $t_{\min} = \Theta(\sqrt{\varepsilon})$. The work of Abbas et al.~\cite{abbas2025nearly} improves the total evolution time to $\mathcal{O}(m/\varepsilon)$, while retaining the same requirement on $t_{\min}$.

Lower bounds for Hamiltonian learning have also been extensively studied~\cite{ma2024learning,abbas2025nearly,chen2025lower,brahmachari2026learning}. 
It is known that any algorithm that learns an $m$-sparse Hamiltonian to accuracy $\varepsilon$ in the Pauli-coefficient $\ell_1$ norm requires a total evolution time of at least $\Omega(m)$~\cite{ma2024learning}. 
This was later strengthened to a lower bound of $\Omega(m/\varepsilon)$ on the total evolution time~\cite{abbas2025nearly}. 
Furthermore, learning a $m$-sparse Hamiltonian $H$ in the $\ell_\infty$ norm requires at least $\Omega(m/n)$ queries (or interactions) with $H$. Naturally, it follows that for $t_{\min}=T$, the total evolution $t_\mathrm{tot}$ is bounded by $\Omega(mT/n)$. It has also been shown that learning a Hamiltonian using at most $\mathcal{L}$ discrete control operations to accuracy $\varepsilon$ in the $\ell_\infty$ norm requires total evolution time $\Omega(1/(\mathcal{L}\varepsilon^{2}))$~\cite{hu2025ansatz}. Consequently, achieving Heisenberg-limited scaling necessitates at least $\Omega(1/\varepsilon)$ discrete control operations.

Practical aspects of Hamiltonian learning in the presence of static single-qubit fields have also been investigated~\cite{brahmachari2026learning}. 
In this setting, achieving Heisenberg-limited scaling in the total evolution time requires either $\mathcal{O}(1/\varepsilon)$ discrete control operations or a static field strength $v=\Omega(1)$. 
Consequently, in the absence of static single-qubit control, one must interact with the Hamiltonian at least $\mathcal{O}(1/\varepsilon)$ times. This implies that only with discrete control, for $t_{\min}=T$, the total evolution $t_\mathrm{tot}$ is bounded by $\Omega(T/\varepsilon)$. 

Finally, recent work on quantum probe tomography~\cite{chen2025probe} provides an alternative framework for Hamiltonian learning under restricted access models. In particular, it has been shown that certain structured Hamiltonians (e.g., translation- and rotation-invariant nearest-neighbor systems) can be reconstructed from local probes of Gibbs states with query complexity polynomial in $1/\varepsilon$ and classical post-processing time polylogarithmic in $1/\varepsilon$. These results demonstrate that Hamiltonian learning remains feasible even under severely constrained measurement access.

\medskip
\noindent
\textbf{Concurrent work.} Concurrently with and independently of this work, Cedillo Vayson de Pradenne, Cotler, and Huang study the problem of learning $U = e^{-iHt}$ for a fixed, large $t$ that cannot be varied. Their results establish that, under mild assumptions on the class of $H$, the learning problem is solvable~\cite{constantin2026}.

\section{Preliminaries}\label{sec:prelim}

We represent any traceless $n$-qubit Hamiltonian $H$ using its Pauli decomposition
\begin{equation}
    H = \sum_x \alpha_x P_x,
\end{equation}
where $\alpha_x \in \mathds{R}$ and $P_x$ ranges over the non-identity Pauli operators.

We use several norms to quantify the size of a Hamiltonian.
The \emph{(unnormalized) Frobenius norm} is defined as
\begin{equation}
    \|H\|_{\mathrm{Frob}} = \sqrt{\operatorname{Tr}(H^\dagger H)}.    
\end{equation}
The \emph{normalized Frobenius norm} is
\begin{equation}
    \|H\|_{F}
    = \sqrt{\frac{\operatorname{Tr}(H^\dagger H)}{2^n}}
    = 2^{-n/2}\,\|H\|_{\mathrm{Frob}}
    = \left( \sum_x |\alpha_x|^2 \right)^{1/2},
\end{equation}
which coincides with the $\ell_2$ norm of the Pauli coefficient vector.

The operator norm is defined as
\begin{equation}
    \|H\|_{\infty} = \|H\|_{\mathrm{op}}
= \sup_{\|v\|=1}\|Hv\|.    
\end{equation}
For Hermitian $H$, this equals the spectral norm $\max_i |\lambda_i|$.
Since all Hamiltonians considered in this work are Hermitian, we use the notation $\|\cdot\|_{\infty}$ throughout.

We will also use the unitary distance
\begin{equation}
    d(U,V) = \min_{\phi}\|U-e^{i\phi}V\|_F .    
\end{equation}
Since $d$ is a metric, it satisfies the triangle inequality.

We also define the Pauli-coefficient $\ell_p$ norm of $H$ as
\begin{equation}
    \|H\|_{\ell_p} = |\mu|_p = \left(\sum_x|\alpha_x|^p\right)^{1/p}.
\end{equation}
In particular, $\|H\|_{\ell_\infty} = \max_x |\alpha_x|$.

We will frequently use the following lemma relating the Pauli-coefficient norms.

\begin{lemma}[Pauli-coefficient growth under powers]\label{lem:Pauli-cofficient-ineq}
Let $A=\sum_x \alpha_x P_x$ be an $m$-sparse Hamiltonian in the Pauli basis.  
Then for every integer $k\ge 1$,
\begin{equation}
    \|A^k\|_{\ell_\infty}\le m^{k-1}\|A\|_{\ell_\infty}^k.    
\end{equation}
\end{lemma}

\begin{proof}
Let $S := \operatorname{supp}(\alpha) = \{x : \alpha_x \neq 0\}$, $|S| \le m$.    

Then
\begin{equation}
    A^k = \sum_{x_1,\dots,x_k \in S}
\alpha_{x_1}\cdots \alpha_{x_k}\, P_{x_1}\cdots P_{x_k}.    
\end{equation}

Fix a Pauli operator $P_y$. The coefficient of $P_y$ in $A^k$ is obtained by summing over all tuples $(x_1,\dots,x_k)\in S^k$ satisfying $P_{x_1}\cdots P_{x_k} = \omega P_y$ for some phase $\omega \in \{\pm 1, \pm i\}$.

Once $x_1,\dots,x_{k-1}$ are fixed, there is at most one choice of $x_k$ such that the above equality holds. Hence the number of contributing tuples is at most $m^{k-1}$. Each term in the sum has magnitude at most $\|A\|_{\ell_\infty}^k$, which implies
\begin{equation}
    |\text{coeff}_{P_y}(A^k)|
\le m^{k-1}\|A\|_{\ell_\infty}^k .    
\end{equation}
Taking the maximum over $y$ yields the desired bound.
\end{proof}

\subsection{Bell sampling}\label{sec:bell-sample}

We recall the Bell basis and the Bell-sampling primitive used throughout this work. For two qubits, the Bell states are
\begin{align}
    \ket{\sigma_{00}}=\frac{1}{\sqrt{2}}(\ket{00}+\ket{11}), \quad
    \ket{\sigma_{01}}=\frac{1}{\sqrt{2}}(\ket{00}-\ket{11}), \quad
    \ket{\sigma_{10}}=\frac{1}{\sqrt{2}}(\ket{01}+\ket{10}), \quad
    \ket{\sigma_{11}}=\frac{1}{\sqrt{2}}(\ket{01}-\ket{10}).    
\end{align}
The $2^n$-qubit Bell basis is given by all tensor products $\ket{\sigma_{\mathbf{s}}}=\ket{\sigma_{s_1}}\otimes\cdots\otimes\ket{\sigma_{s_n}}$, where $s_i\in\{00,01,10,11\}$. Equivalently, we may index outcomes by $\mathbf{s}\in\{0,1\}^{2n}$ via the standard identification
$\{00,01,10,11\}\cong\{0,1\}^2$ per qubit-pair.

Given an $n$-qubit state $\rho$, \emph{Bell sampling}~\cite{montanaro2017learning,hangleiter2024bell}
means measuring $\rho^{\otimes 2}$ in the $n$-fold Bell basis
$\{\ket{\sigma_{\mathbf{s}}}\}_{\mathbf{s}\in\{0,1\}^{2n}}$.
The resulting outcome $\mathbf{s}$ is distributed according to the \emph{characteristic distribution}
$q_\rho(\mathbf{s})$.
For a pure state $\rho=\ket{\psi}\!\bra{\psi}$, this distribution admits the Pauli-moment form
\begin{equation}\label{eq:characteristic-distribution}
q_\rho(\mathbf{s})
=2^{-n}\,\bigl\langle \psi \big| P_{\mathbf{s}} \big| \psi \bigr\rangle^{2},
\end{equation}
where $P_{\mathbf{s}}$ denotes the $n$-qubit Pauli operator associated with~$\mathbf{s}$.
Thus, Bell sampling provides direct access to squared Pauli expectation values of the underlying state.

A useful consequence of~\cref{eq:characteristic-distribution} is the following.
If a unitary admits a Pauli expansion $U=\sum_x U_x\,\sigma_x$, then Bell sampling applied to the Choi state $(U\otimes I)\ket{\Omega}$
returns the label $x$ with probability $|U_x|^2$.
Equivalently, there exists a fixed unitary $V$ such that
\begin{equation}
    V(U\otimes I)\ket{\Omega}=\sum_x U_x\ket{x},
\end{equation}
where $V$ transforms $(\sigma_x\otimes I)\ket{\Omega}$ into $\ket{x}$.

\subsection{Pure-state tomography}

Our Hamiltonian learning algorithm reduces the problem to sparse pure-state tomography. We therefore analyze the sample complexity of reconstructing a sparse pure quantum state under two different norms. In particular, we provide guarantees for recovery in the $\ell_\infty$ and $\ell_2$ norms of the coefficient vector.

\begin{lemma}[Sparse pure-state tomography in $\ell_\infty$ norm]
\label{lem:sparse-tomo-linf}
Let $\ket{\psi}=\sum_x \alpha_x\ket{x}$ be a pure quantum state, and let $\varepsilon,\delta\in(0,1)$.
Assume there exists a set $S_\star$ with $|S_\star|\le s$ such that $|\alpha_0-1|\le {\varepsilon}/{3}$ and $|\alpha_x|\le {\varepsilon}/{3}$ for all $x\notin S_\star$. Then there exists a randomized procedure that, using
\begin{equation}
    \mathcal{O}\!\left(\frac{s}{\varepsilon^2}\log\frac{s}{\delta}\right)
\end{equation}
copies of $\ket{\psi}$, outputs a coefficient vector $\widetilde{\alpha}$ satisfying $|\widetilde{\alpha}-\alpha|_\infty \le \varepsilon$ with probability at least $1-\delta$.
\end{lemma}

\begin{proof}
First measure $\mathcal{O}({\varepsilon^{-2}}\log({s}/{\delta}))$ copies of $\ket{\psi}$ in the computational basis and keep all indices whose empirical probability exceeds a suitable constant multiple of $\varepsilon^2$.

By standard heavy-hitters guarantees, with probability at least $1-\delta/3$, the procedure returns a set $T$ such that $|\alpha_x|>{3\varepsilon}/{4}$ implies $x\in T$ and $|\alpha_x|<{3\varepsilon}/{8}$ implies $x\notin T$. Since $|\alpha_x|\le \varepsilon/3<3\varepsilon/8$ for $x\notin S_\star$, we have $T\subseteq S_\star$ and $|T|\le s$. 

Moreover, every $x\notin T$ satisfies $|\alpha_x|\le 3\varepsilon/4$, and hence $|\alpha_{T^c}|_\infty \le {3\varepsilon}/{4}$. Also, $|\alpha_0|\ge 1-\varepsilon/3>3/4$, so $0\in T$.

Next perform standard pure-state tomography restricted to the support $T$ to accuracy $\varepsilon/20$.
Using
\begin{equation}
    \mathcal{O}\!\left(\frac{|T|}{\varepsilon^2}\log\frac{|T|}{\delta}\right)
    =
    \mathcal{O}\!\left(\frac{s}{\varepsilon^2}\log\frac{s}{\delta}\right)
\end{equation}
copies, we obtain coefficients $\widehat{\alpha}$ supported on $T$ such that, with probability at least $1-\delta/3$,
\begin{equation}
    \left\|\widehat{\alpha}-e^{i\phi}\alpha_T\right\|_2\le \frac{\varepsilon}{20}
\end{equation}
for some $\phi\in\mathbb{R}$.

Define
\begin{equation}
    \widetilde{\alpha}_x :=
    \begin{cases}
        \overline{\widehat{\alpha}_0}\,\widehat{\alpha}_x, & x\in T,\\
        0, & x\notin T.
    \end{cases}
\end{equation}
Since
\begin{align}
    |\widehat{\alpha}_0-e^{i\phi}|
    & \le
    |\widehat{\alpha}_0-e^{i\phi}\alpha_0|+|\alpha_0-1| \\
    & \le
    \frac{\varepsilon}{20}+\frac{\varepsilon}{3},
\end{align}
the phase correction ensures $\|\widetilde{\alpha}_T-\alpha_T\|_\infty \le {\varepsilon}/{2}$. Therefore $\|\widetilde{\alpha}-\alpha\|_\infty
    \le
    \max\!\left\{
        \|\widetilde{\alpha}_T-\alpha_T\|_\infty,\,
        \|\alpha_{T^c}\|_\infty
    \right\}
    \le \varepsilon$.

The claimed sample complexity follows by summing the two stages, and the failure probability is at most $\delta$ by a union bound.
\end{proof}

\begin{lemma}[Sparse pure-state tomography in $\ell_2$ norm]
\label{lem:sparse-tomo-l2}
Let $\ket{\psi}=\sum_x \alpha_x\ket{x}$ be a pure quantum state, and let $\varepsilon,\delta\in(0,1)$.
Assume there exists a set $S_\star$ with $|S_\star|\le s$ such that $|\alpha_0-1|\le {\varepsilon}/{4}$ and $\left(\sum_{x\notin S_\star}|\alpha_x|^2\right)^{1/2}\le {\varepsilon}/{4}$.
Then there exists a randomized procedure that, using
\begin{equation}
    \mathcal{O}\!\left(\frac{s}{\varepsilon^2}\log\frac{s}{\delta}\right)
\end{equation}
copies of $\ket{\psi}$, outputs a coefficient vector $\widetilde{\alpha}$ satisfying $|\widetilde{\alpha}-\alpha|_2 \le \varepsilon$ with probability at least $1-\delta$.
\end{lemma}

\begin{proof}
First measure $\mathcal{O}({s}{\varepsilon^{-2}}\log({s}/{\delta}))$ copies of $\ket{\psi}$ in the computational basis and keep all indices whose empirical probability exceeds a suitable constant multiple of $\varepsilon^2/s$.

By standard heavy-hitters guarantees, with probability at least $1-\delta/3$, the procedure produces a set $T$ such that $|\alpha_x|>{\varepsilon}/{(2\sqrt{s})}$ implies $x\in T$ and $|\alpha_x|<{\varepsilon}/{(4\sqrt{s})}$ implies $x\notin T$, and hence $|T|=O(s)$.

Moreover,
\begin{align}
    \|\alpha_{T^c}\|_2
    &\le
    \|\alpha_{S_\star^c}\|_2
    +
    \|\alpha_{S_\star\cap T^c}\|_2 \\
    & \le
    \frac{\varepsilon}{4}
    +
    \sqrt{s}\cdot\frac{\varepsilon}{2\sqrt{s}} \\
    & =
    \frac{3\varepsilon}{4}.
\end{align}

Also, $|\alpha_0|\ge 1-\varepsilon/4>3/4$, so $0\in T$.

Next perform standard pure-state tomography restricted to the support $T$ to accuracy $\varepsilon/20$.
Since $|T|=O(s)$, this uses
\begin{equation}
    \mathcal{O}\!\left(\frac{|T|}{\varepsilon^2}\log\frac{|T|}{\delta}\right)
    =
    \mathcal{O}\!\left(\frac{s}{\varepsilon^2}\log\frac{s}{\delta}\right)
\end{equation}
copies and returns coefficients $\widehat{\alpha}$ supported on $T$ such that, with probability at least $1-\delta/3$,
\begin{equation}
    \left\|\widehat{\alpha}-e^{i\phi}\alpha_T\right\|_2\le \frac{\varepsilon}{20}
\end{equation}
for some $\phi\in\mathbb{R}$.

Define
\begin{equation}
    \widetilde{\alpha}_x :=
    \begin{cases}
        \overline{\widehat{\alpha}_0}\,\widehat{\alpha}_x, & x\in T,\\
        0, & x\notin T.
    \end{cases}
\end{equation}

Since
\begin{align}
    \left|\widehat{\alpha}_0-e^{i\phi}\right|
    & \le
    \left|\widehat{\alpha}_0-e^{i\phi}\alpha_0\right|
    +
    |\alpha_0-1| \\
    & \le
    \frac{\varepsilon}{20}
    +
    \frac{\varepsilon}{4},
\end{align}
the same phase-correction argument yields $\|\widetilde{\alpha}_T-\alpha_T\|_2 \le {\varepsilon}/{4}$. Therefore
\begin{align}
    \|\widetilde{\alpha}-\alpha\|_2
    & \le
    \|\widetilde{\alpha}_T-\alpha_T\|_2
    +
    \|\alpha_{T^c}\|_2 \\
    & \le
    \frac{\varepsilon}{4}
    +
    \frac{3\varepsilon}{4} \\
    & =
    \varepsilon.
\end{align}

The claimed sample complexity follows by summing the two stages, and the failure probability is at most $\delta$ by a union bound.
\end{proof}

\subsection{Technical tools}
In this section, we collect several useful mathematical lemmas that will be used throughout the proofs. We begin with Duhamel's principle, which provides a convenient way to express and bound $e^{-iXt}e^{iYt}$.

\begin{lemma}[Duhamel's principle~\cite{mathias1992evaluating}]\label{lem:duhamel}
Let $X,Y$ be Hermitian matrices, and let $V(t):=e^{-iXt}e^{iYt}$ and $\Delta:=X-Y$.
Then, for every $t\ge 0$,
\begin{equation}
    V(t)=I-i\int_0^t e^{-iXs}\,\Delta\,e^{iYs}\,ds.
\end{equation}
Consequently, for any unitarily invariant norm $\|\cdot\|$, we have $\|V(t)-I\| \le t\|\Delta\|$.
\end{lemma}



We next show that any unitary can be written, up to a global phase, as the exponential of a traceless Hermitian matrix with a controlled norm.

\begin{lemma}[Traceless Hermitian representation of a unitary]\label{lem:log}
Let $U$ be a unitary matrix. Then there exists a traceless Hermitian matrix $H$ such that $d(U,e^{-iH})=0$, where $d(U,V):=\min_{\phi\in\mathbb{R}} |U-e^{i\phi}V|$ and $\|\cdot\|$ is any unitarily invariant norm. Moreover, $\|H\| \le \pi \|I-U\|$.
\end{lemma}

\begin{proof}
By the spectral theorem, there exists an orthonormal basis
$\{\ket{\psi_j}\}_{j=1}^d$ and phases $\theta_j \in (-\pi,\pi]$ such that
\begin{equation}
    U = \sum_{j=1}^d e^{-i\theta_j}\ket{\psi_j}\bra{\psi_j}.
\end{equation}
Define
\begin{align}
    \bar{\theta} &:= \frac{1}{d}\sum_{j=1}^d \theta_j, \\
    H &:= \sum_{j=1}^d (\theta_j-\bar{\theta})\ket{\psi_j}\bra{\psi_j}.
\end{align}
Then $H$ is Hermitian and traceless, since
\begin{equation}
    \Tr(H)=\sum_{j=1}^d (\theta_j-\bar{\theta})=0.
\end{equation}

Moreover,
\begin{equation}
    e^{-iH}
    = \sum_{j=1}^d e^{-i(\theta_j-\bar{\theta})}\ket{\psi_j}\bra{\psi_j}
    = e^{i\bar{\theta}} U,
\end{equation}
and hence $d(U,e^{-iH})=0$.

We now prove the norm bound. Since $\|\cdot\|$ is unitarily invariant,
it depends only on the singular values. In the eigenbasis of $U$, both $H$ and $I-U$ are diagonal:
\begin{align}
    H &= \mathrm{diag}(\theta_1-\bar{\theta},\dots,\theta_d-\bar{\theta}), \\
    I-U &= \mathrm{diag}(1-e^{-i\theta_1},\dots,1-e^{-i\theta_d}).
\end{align}

For each $j$, we have
\begin{equation}
    |1-e^{-i\theta_j}| = 2\left|\sin\frac{\theta_j}{2}\right|
    \ge \frac{2|\theta_j|}{\pi},
\end{equation}
and hence
\begin{equation}
    |\theta_j| \le \frac{\pi|1-e^{-i\theta_j}|}{2}.
\end{equation}
Let $M := \max_j |\theta_j|$. Then $|\bar{\theta}|\le M$, and therefore
\begin{equation}
    |\theta_j-\bar{\theta}|
    \le |\theta_j| + |\bar{\theta}|
    \le 2M
    \le \pi \max_k \left|1-e^{-i\theta_k}\right|.
\end{equation}
Since $\Phi$ is monotone with respect to coordinate-wise inequalities, it follows that $\|H\| \le \pi \, \|I-U\|$.
\end{proof}

We next invoke the Baker--Campbell--Hausdorff (BCH) formula to express the product of exponentials as a single exponential, enabling a term-by-term analysis of the resulting generator.

\begin{lemma}[Baker--Campbell--Hausdorff formula, Dynkin form~\cite{dynkin1949representation}]
\label{lem:BCH}
Let $X,Y$ be bounded operators such that $\|X\|_\infty + \|Y\|_\infty < \log 2$. 
Then the Baker--Campbell--Hausdorff series converges absolutely, and
\begin{equation}
    \log(e^X e^Y)=\sum_{r\ge 1}\mathrm{BCH}_r(X,Y).
\end{equation}
Here $\mathrm{BCH}_r(X,Y)$ is the homogeneous Lie polynomial of degree $r$ given by
\begin{equation}
\mathrm{BCH}_r(X,Y)
=
\sum_{n=1}^r \frac{(-1)^{n-1}}{n}
\sum_{\substack{r_i,s_i\ge 0\\ r_i+s_i>0\\ \sum_{i=1}^n(r_i+s_i)=r}}
\frac{
[X^{r_1}Y^{s_1}\cdots X^{r_n}Y^{s_n}]
}{
r\prod_{i=1}^n (r_i!\,s_i!)
}.
\end{equation}
Here $[X^{r_1}Y^{s_1}\cdots X^{r_n}Y^{s_n}]$ denotes the left-normed commutator associated with the word consisting of $r_1$ copies of $X$, followed by $s_1$ copies of $Y$, and so on. 
Equivalently, whenever $s_n \ge 1$, we have
\begin{equation}
[X^{r_1}Y^{s_1}\cdots X^{r_n}Y^{s_n}]
=
(\operatorname{ad}_X)^{r_1}(\operatorname{ad}_Y)^{s_1}
\cdots
(\operatorname{ad}_X)^{r_n}(\operatorname{ad}_Y)^{s_n-1}(Y),
\end{equation}
where $\operatorname{ad}_X(Z) := [X,Z]$, and $\operatorname{ad}_X^k$ denotes its $k$-fold iteration.
\end{lemma}

We next derive a bound on each BCH term in the normalized Frobenius norm, which will be used to control the contribution of higher-order terms.

\begin{corollary}[Degreewise normalized Frobenius bound for BCH coefficients]
\label{cor:BCH_F_bound}
Assume the hypotheses of Lemma~\ref{lem:BCH}, and let $M:=\max\{\|X\|_\infty, \|Y\|_\infty\}$ and $\varepsilon:=\|X+Y\|_F$. Then 
\begin{equation} 
\|\mathrm{BCH}_r(X,Y)\|_F \le C_r\, M^{r-1}\varepsilon,
\end{equation}
where $C_r:=2^{2r-1}{r^r}/{r!} \le (4e)^r$.
\end{corollary}

\begin{proof}
Define the left-normed commutator $L_r(Z_1,\dots,Z_r)
:=
[Z_1,[Z_2,\dots,[Z_{r-1},Z_r]\dots]]$. We will use the inequality $\|[U,V]\|_F\le 2\|U\|_F\|V\|_\infty$. 

By induction on $r$, it follows that for every $1\le j\le r$,
\begin{equation}\label{eq:left-normed-bound-plus}
\|L_r(Z_1,\dots,Z_r)\|_F
\le
2^{r-1}\|Z_j\|_F\prod_{i\ne j}\|Z_i\|_\infty.
\end{equation}

Now fix $r\ge 2$ (the case $r=1$ is trivial), $n\in\{1,\dots,r\}$, and a tuple $(r_1,s_1,\dots,r_n,s_n)$ appearing in Lemma~\ref{lem:BCH}. Let $W=[X^{r_1}Y^{s_1}\cdots X^{r_n}Y^{s_n}]
= L_r(Z_1,\dots,Z_r)$, where $Z_j\in\{X,Y\}$.

Define the reference tuple \((\widetilde Z_1,\dots,\widetilde Z_r)\) by
\begin{equation}
    \widetilde Z_j=
\begin{cases}
X, & Z_j=X,\\
-\,X, & Z_j=Y.
\end{cases}    
\end{equation}
Since each $\widetilde Z_j$ is a scalar multiple of $X$, we have $L_r(\widetilde Z_1,\dots,\widetilde Z_r)=0$ for $r\ge 2$. By multilinearity and telescoping,
\begin{equation}
    L_r(Z_1,\dots,Z_r)
=
\sum_{j=1}^r
L_r \left(\widetilde Z_1,\dots,\widetilde Z_{j-1},\,Z_j-\widetilde Z_j,\,Z_{j+1},\dots,Z_r\right).
\end{equation}

If $Z_j=X$, then $Z_j-\widetilde Z_j=0$, while if $Z_j=Y$, then $Z_j-\widetilde Z_j = X+Y$. Hence only positions with $Z_j=Y$ contribute. Let $q$ be the number of occurrences of $Y$, so $q\le r$. Using \Cref{eq:left-normed-bound-plus}, we obtain
\begin{equation}\label{eq:single-word-bound-plus}    
    \|W\|_F
\le
q\,2^{r-1}M^{r-1}\|X+Y\|_F
\le
r\,2^{r-1}M^{r-1}\varepsilon.
\end{equation}

Applying \cref{eq:single-word-bound-plus} to the Dynkin formula in Lemma~\ref{lem:BCH}, the factor $r$ cancels the denominator $r$, yielding
\begin{equation}\label{eq:bch-pre-comb-plus}
    \|\mathrm{BCH}_r(X,Y)\|_F \le
\left(2^{r-1}M^{r-1}\varepsilon \right) \cdot
\sum_{n=1}^r \left( \frac{1}{n}
\sum_{\substack{r_i,s_i\ge 0\\ r_i+s_i>0\\ \sum_{i=1}^n(r_i+s_i)=r}}
\frac{1}{\prod_{i=1}^n r_i!\,s_i!} \right). 
\end{equation}

To bound the inner sum, set $d_i:=r_i+s_i\ge 1$. Then
\begin{equation}
    \sum_{r_i+s_i=d_i}\frac{1}{r_i!\,s_i!}
=
\frac{2^{d_i}}{d_i!},    
\end{equation}
and hence
\begin{equation}
    \sum_{\substack{r_i,s_i\ge 0\\ r_i+s_i>0\\ \sum_{i=1}^n(r_i+s_i)=r}}
\frac{1}{\prod_{i=1}^n r_i!\,s_i!}
=
2^r
\sum_{\substack{d_1,\dots,d_n\ge 1\\ d_1+\cdots+d_n=r}}
\frac{1}{d_1!\cdots d_n!}.    
\end{equation}

Define
\begin{equation}
    T_{r,n}
:=
\sum_{\substack{d_1,\dots,d_n\ge 1\\ d_1+\cdots+d_n=r}}
\frac{1}{d_1!\cdots d_n!}.    
\end{equation}
Then $(e^t-1)^n=\sum_{r\ge n}T_{r,n}t^r$, so
\begin{equation}
    T_{r,n}=[t^r](e^t-1)^n=\frac{n!\,S(r,n)}{r!},    
\end{equation}
where $S(r,n)$ is a Stirling number of the second kind. Since $n!\,S(r,n)\le n^r$, we obtain $T_{r,n}\le {n^r}/{r!}$.

Substituting into \Cref{eq:bch-pre-comb-plus}, we obtain
\begin{align}
    \|\mathrm{BCH}_r(X,Y)\|_F &\le
\left(2^{2r-1}M^{r-1}\varepsilon \right) \cdot
\sum_{n=1}^r \frac{1}{n}T_{r,n} \\
& \le
\left( 2^{2r-1}M^{r-1}\varepsilon \right) \cdot \frac{1}{r!}\sum_{n=1}^r n^{r-1}.    
\end{align}
Using $\sum_{n=1}^r n^{r-1}\le r^r$, we conclude
\begin{equation}
    \|\mathrm{BCH}_r(X,Y)\|_F \le \frac{2^{2r-1}r^r}{r!}M^{r-1}\varepsilon,
\end{equation}
for $r \ge 2$.

Finally, Stirling's bound $r!\ge \sqrt{2\pi r}(r/e)^r$ implies
\begin{equation}
    C_r = \frac{2^{2r-1}r^r}{r!} \le \frac{2^{2r-1}e^r}{\sqrt{2\pi r}} = \frac{(4e)^r}{2\sqrt{2\pi r}} \le (4e)^r,    
\end{equation}
completing the proof.
\end{proof}

We next introduce a projection step onto the set of sparse, norm-bounded Hamiltonians, which will be used to maintain feasibility during the iteration.

\begin{definition}[Sparse bounded truncation]\label{lem:sparse-bound-trunc}
Let $\mathcal{S}_{k,c} := \left\{ H_0 : \operatorname{supp}_P(H_0) \le k,\ \|H_0\|_{\infty} \le c \right\}$ be the set of $k$-sparse Hamiltonians with operator norm at most $c$. For a Hamiltonian $H$, define
\begin{equation}
    \mathcal{T}_{k,c}(H) := \operatorname*{arg\,min}_{H_0 \in \mathcal{S}_{k,c}} \|H - H_0\|_{\ell_\infty}.    
\end{equation}
\end{definition}

\begin{lemma}[Stability of sparse bounded truncation]\label{lem:sbt-stab}
    Let $A$ be an $m$-sparse Hamiltonian with $\|A\|_{\infty} \le c$, and let $B$ be a Hamiltonian satisfying $\|A-B\|_{\ell_\infty} \le \varepsilon$. Then $\|A-\mathcal{T}_{m,c}(B)\|_{\ell_\infty} \le 2\varepsilon$.
\end{lemma}
\begin{proof}
Since $A \in \mathcal{S}_{m,c}$, optimality gives
$\|B-\mathcal{T}_{m,c}(B)\|_{\ell_\infty} \le \|B-A\|_{\ell_\infty} \le \varepsilon$.
The claim follows by the triangle inequality.
\end{proof}

\section{Reduction to pure-state tomography}\label{sec:repure}

\subsection{Correctness of Algorithm~\ref{alg:pure-l-infty}}

In this section, we reduce Hamiltonian learning to pure-state tomography. 
For convenience, define a unitary $V$ satisfying $V(P_x\otimes I)\ket{\Phi}=\ket{x}$ for all $x$. 
Such a unitary exists since the states $\{(P_x\otimes I)\ket{\Phi} : x\in\{0,1\}^{2n}\}$ form an orthonormal basis.

\begin{theorem}[Sparse Hamiltonian learning via pure-state tomography]\label{thm:ham-learning-tomo}
Let $A=\sum_{x\in\{0,1\}^{2n}} \alpha_x P_x$ be an $n$-qubit traceless Hermitian operator, where $\{P_x\}$ is the Pauli basis and $|\operatorname{supp}(\alpha)|\le m$. 
Assume that $\|A\|_{\ell_\infty}\le \varepsilon$. 
Then, using queries to the real-time evolution operator $e^{-iAt}$ with $t=\mathcal{O}(1/(m\varepsilon))$, we can learn an $m$-sparse Hamiltonian $A_1$ such that $\|A-A_1\|_{\ell_\infty} \le \varepsilon/8$. 
The total evolution time is $\mathcal{O}(m^2/\varepsilon)$.
\end{theorem}

To prove the~\Cref{thm:ham-learning-tomo}, we first introduce the following lemma.

\begin{lemma}[First-order reduction to pure-state tomography]\label{lem:ham-to-tomo}
Let $A=\sum_x \alpha_x P_x$ be an $m$-sparse traceless Hamiltonian with $\|A\|_{\ell_\infty} \le \varepsilon$. Define
\begin{equation}
    \ket{\psi_t} = V(e^{-iAt}\otimes I)\ket{\Phi} = \sum_x \beta_x \ket{x},    
\end{equation}
where $t \le 1/(m\varepsilon)$. Then $|1-\beta_0| \le mt^2\varepsilon^2$, and $|\beta_x + it\alpha_x| \le mt^2\varepsilon^2$ for all $x \ne 0$. In particular, learning $\ket{\psi_t}$ allows one to recover the coefficients $\alpha_x$.
\end{lemma}

\begin{proof}
Let
\begin{equation}
    R := e^{-iAt} - (I - iAt) = \sum_{k \ge 2} \frac{(-itA)^k}{k!}.    
\end{equation}

Then
\begin{align}
\|R\|_{\ell_\infty}
&\le \sum_{k \ge 2} \frac{t^k}{k!} \|A^k\|_{\ell_\infty} \\
&\le \sum_{k \ge 2} \frac{t^k}{k!} m^{k-1}\varepsilon^k \\
&= mt^2\varepsilon^2 \sum_{k \ge 2} \frac{(tm\varepsilon)^{k-2}}{k!}\\
& \le mt^2\varepsilon^2 \sum_{k \ge 2} \frac{1}{k!} \\
& < mt^2\varepsilon^2,
\end{align}
where we used Lemma~\ref{lem:Pauli-cofficient-ineq}.

Since $\alpha_0 = \Tr(A) = 0$, we have
\begin{equation}
    V((I - iAt)\otimes I)\ket{\Phi}
= \ket{0} - it \sum_{x \ne 0} \alpha_x \ket{x}.    
\end{equation}
On the other hand,
\begin{equation}
    V(e^{-iAt}\otimes I)\ket{\Phi}
= \sum_x \beta_x \ket{x}.    
\end{equation}

Since the Pauli coefficients of $R$ are bounded by $\|R\|_{\ell_\infty}$, we have $|\beta_0 - 1| \le mt^2\varepsilon^2$ and $|\beta_x + it\alpha_x| \le mt^2\varepsilon^2$ for all $x \ne 0$.
\end{proof}

Now we are ready to prove Theorem~\ref{thm:ham-learning-tomo}.
\begin{proof}[Proof of~\Cref{thm:ham-learning-tomo}]
    Since $A$ is $m$-sparse, there exists a set $S_\star$ with $|S_\star|=m$ such that $\alpha_x=0$ for all $x\notin S_\star$ and $0\notin S_\star$.     

    By Lemma~\ref{lem:ham-to-tomo}, we have $|\beta_0-1|\le mt^2\varepsilon^2$ and $|\beta_x|\le mt^2\varepsilon^2$ for all $x\notin S_\star$. 
Applying Lemma~\ref{lem:sparse-tomo-linf}, we obtain an estimate $\widetilde{\beta}=\{\widetilde{\beta}_x\}$ satisfying $|\widetilde{\beta}_x-\beta_x|\le 3mt^2\varepsilon^2$ using $\widetilde{\mathcal{O}}(1/(mt^4\varepsilon^4))$ samples of $\ket{\psi_t}$.

Since $\alpha_x\in\mathbb{R}$, for $x\ne 0$ define $\widetilde{\alpha}_x:=-\operatorname{Im}(\widetilde{\beta}_x)/t$. Then
\begin{align}
\left|\widetilde{\alpha}_x-\alpha_x\right|
&\le \frac{|\widetilde{\beta}_x-\beta_x| + |\beta_x + it\alpha_x|}{t} \\
&\le \frac{3mt^2\varepsilon^2 + mt^2\varepsilon^2}{t} \\
& = 4mt\varepsilon^2.
\end{align}

Define $\widetilde{A}:=\sum_x \widetilde{\alpha}_x P_x$. Then $\widetilde{A}$ is Hermitian. 
Setting $t=1/(32m\varepsilon)$ yields
\begin{equation}
    \|A-\widetilde{A}\|_{\ell_\infty} \le 4mt\varepsilon^2 \le \frac{\varepsilon}{8}.    
\end{equation}

The number of samples is $\widetilde{\mathcal{O}}(1/(mt^4\varepsilon^4)) = \widetilde{\mathcal{O}}(m^3)$, and hence the total evolution time is $\widetilde{\mathcal{O}}(m^3 t) = \widetilde{\mathcal{O}}(m^2/\varepsilon)$.
\end{proof}

\subsection{Correctness of Algorithm~\ref{alg:pure-Frobenius}}
The correctness of Algorithm~\ref{alg:pure-Frobenius} is guaranteed provided that $\varepsilon \le {c}/{(10c_F c_\infty)}$.

\begin{theorem}[Sparse Hamiltonian learning in Frobenius norm via pure-state tomography]\label{thm:ham-learning-tomo-F}
Let $A=\sum_{x\in\{0,1\}^{2n}} \alpha_x P_x$ be an $n$-qubit traceless Hermitian operator, where $\{P_x\}$ is the Pauli basis and $|\operatorname{supp}_P(A, c\varepsilon/8)|\le s$. Assume that $\|A\|_{\infty}\le c_{\infty}\varepsilon$ and $\|A\|_F\le c_F\varepsilon$. Then, using queries to the real-time evolution operator $e^{-iAt}$ with $t=\mathcal{O}(c/(c_Fc_\infty\varepsilon))$, we can learn an $s$-sparse Hamiltonian $A_1$ such that $\|A-A_1\|_F \le c\varepsilon$. The total evolution time is 
\begin{equation}
    \widetilde{\mathcal{O}}\!\left(\frac{s c_F c_\infty}{c^3 \varepsilon}\right).
\end{equation}
\end{theorem}

\begin{proof}
Let $R := e^{-iAt} - (I - iAt)$. Then $\|R\|_F \le t^2 \|A\|_F \|A\|_{\infty} \le c_F c_\infty t^2 \varepsilon^2$, provided that $\|A\|_{\infty} t \le 1$, which holds for $t \le 1/(c_\infty \varepsilon)$.

Define $\ket{\psi_t} = V(e^{-iAt}\otimes I)\ket{\Phi} = \sum_x \beta_x \ket{x}$. We also have,
\begin{equation}
    V((I - iAt)\otimes I)\ket{\Phi}
= \ket{0} - it \sum_{x \ne 0} \alpha_x \ket{x}.    
\end{equation}
Since the Pauli coefficients of $R$ are bounded by $\|R\|_F$, we have $|\beta_0 - 1| \le c_F c_\infty t^2 \varepsilon^2$ and
\begin{equation}
    \left(\sum_{x \ne 0} |\beta_x + it\alpha_x|^2 \right)^{1/2}
\le c_F c_\infty t^2 \varepsilon^2,
\end{equation}
for $x\neq 0$.

Since $s = \operatorname{supp}_P(A, c\varepsilon/8)$, there exists a set $S_\star$ with $|S_\star|=s$ and $0 \notin S_\star$ such that
\begin{equation}
    \left(\sum_{x \notin S_\star} |\alpha_x|^2 \right)^{1/2} \le \frac{1}{8} c \varepsilon.    
\end{equation}
It follows that $|\beta_0 - 1| \le c_F c_\infty t^2 \varepsilon^2$ and
\begin{equation}
    \left(\sum_{x \notin S_\star} |\beta_x|^2 \right)^{1/2}
\le c_F c_\infty t^2 \varepsilon^2 + \frac{1}{8} c t\varepsilon.    
\end{equation}

Applying Lemma~\ref{lem:sparse-tomo-l2}, we obtain $\widetilde{\beta}$ satisfying $\|\widetilde{\beta} - \beta\|_2
\le 4 c_F c_\infty t^2 \varepsilon^2 + c t \varepsilon / 2$.

Define $\widetilde{\alpha}_x := -\operatorname{Im}(\widetilde{\beta}_x)/t$. Then
\begin{align}
\|\widetilde{\alpha} - \alpha\|_2
&\le \frac{\|\widetilde{\beta} - \beta\|_2 + \|\beta + it\alpha\|_2}{t} \\
&\le \frac{4 c_F c_\infty t^2 \varepsilon^2 +  c t \varepsilon /2 + c_F c_\infty t^2 \varepsilon^2}{t} \\
&= 5 c_F c_\infty t \varepsilon^2 +  c \varepsilon / 2.
\end{align}

Define $\widetilde{A} := \sum_x \widetilde{\alpha}_x P_x$, which is Hermitian. 
Setting $t = \lfloor {c}/{(10 c_F c_\infty \varepsilon)} \rfloor$ yields $\|A - \widetilde{A}\|_F \le c \varepsilon$, provided that ${c}/{(10 c_F c_\infty \varepsilon)} \ge 1$.

The total evolution time is
\begin{equation}
    \widetilde{\mathcal{O}}\!\left(\frac{s t}{(c_F c_\infty t^2 \varepsilon^2)^2}\right)
= \widetilde{\mathcal{O}}\!\left(\frac{s c_F c_\infty}{c^3 \varepsilon}\right).    
\end{equation}
\end{proof}

\section{Approximating continuous quantum control using long-time evolution}\label{sec:conquantumlong}
In this section, we show how to approximate continuous quantum control using only long-time evolution. 

Define $C_j := e^{iHT}e^{-iH_jT}$. By Lemma~\ref{lem:log}, there exists a Hermitian matrix $W_j$ such that $d(C_j, e^{-iW_j}) = 0$. We assume that $H$ and $H_j$ are $m$-sparse traceless Hamiltonians satisfying $\|H - H_j\|_{\ell_\infty} \le \varepsilon$. 
Our goal is to learn a Hermitian matrix $\widetilde{W}$ such that $\|W - \widetilde{W}\|_F \le c \varepsilon$, for a desired constant $c$, using queries to $e^{-iHT}$ and with total evolution time achieving Heisenberg-limited scaling.

This procedure will serve as a subroutine for Hamiltonian learning.

\subsection{Approximating continuous quantum control with arbitrary long time evolution}\label{subsec:arblong}

To efficiently learn the Hamiltonian $W_j$, we first bound its norm and sparsity. 
Using Lemma~\ref{lem:duhamel} and Lemma~\ref{lem:log}, we have
\begin{equation}
    \|W_j\|_F \le \pi \|I-C_j\|_F \le \pi T \|H-H_j\|_F \le 2\pi T \sqrt{m}\,\varepsilon,    
\end{equation}
and
\begin{equation}
    \|W_j\|_{\infty} \le \pi \|I-C_j\|_\infty \le \pi T \|H-H_j\|_\infty \le 2\pi T m \varepsilon,    
\end{equation}
since $\|H-H_j\|_{\ell_\infty} \le \varepsilon$ and both $H,H_j$ are $m$-sparse.

We now bound the sparsity of $W_j$.

\begin{lemma}[Pauli sparsity of a bounded logarithm]\label{lem:sparsity-upper}
Let $H=\sum_{a\in S_H}\alpha_a P_a$ and $H_j=\sum_{b\in S_{H_j}}\beta_b P_b$ be $m$-sparse Hamiltonians. Then any Hermitian $W_j$ satisfying $d(e^{iHT}e^{-iH_jT},e^{-iW_j})=0$ and $\|W_j\|_{\infty} < \pi$ obeys $|\operatorname{supp}_P(W_j)| \le 4^m$.
\end{lemma}

\begin{proof}
Let $L := \operatorname{span}_{\mathbb F_2}(S_H \cup S_{H_j}) \subseteq \mathbb F_2^{2n}$ and $r := \dim L$. Then $r \le 2m$ and $|L| = 2^r$.

Let $\mathcal A_L := \operatorname{span}_{\mathbb C}\{P_v : v \in L\}$. Since Pauli multiplication is closed modulo phase, $\mathcal A_L$ is a finite-dimensional unital $*$-subalgebra. As $H,H_j \in \mathcal A_L$, it follows from the power-series expansion that $e^{iHT}, e^{-iH_jT} \in \mathcal A_L$, and hence
\[
C_j := e^{iHT}e^{-iH_jT} \in \mathcal A_L.
\]

By the assumption $d(C_j,e^{-iW_j})=0$, there exists a phase $\phi$ such that
\[
e^{-iW_j}=e^{i\phi}C_j.
\]
Set $C_{j,\phi}:=e^{i\phi}C_j$. Then $C_{j,\phi}\in\mathcal A_L$ is unitary. Write its spectral decomposition as $C_{j,\phi}=\sum_k \lambda_k \Pi_k$, where the $\lambda_k$'s are distinct and $|\lambda_k|=1$. Since $C_{j,\phi}$ has a finite spectrum, each spectral projector $\Pi_k$ can be expressed as a polynomial in $C_{j,\phi}$, e.g., via Lagrange interpolation, and hence $\Pi_k \in \mathcal A_L$.

It remains to show that the given $W_j$ lies in $\mathcal A_L$. Since $\|W_j\|_\infty<\pi$, the spectrum of $W_j$ lies in $[-1,1]$, where the map $\theta\mapsto e^{-i\theta}$ is injective. Therefore, for each eigenvalue $\lambda_k$ of $C_{j,\phi}=e^{-iW_j}$, there is a unique $\theta_k\in[-1,1]$ such that $\lambda_k=e^{-i\theta_k}$. Hence $W_j$ is scalar on each eigenspace $\operatorname{ran}(\Pi_k)$, and we can write
\[
W_j=\sum_k \theta_k \Pi_k.
\]
Since each $\Pi_k\in\mathcal A_L$, we obtain $W_j\in\mathcal A_L$.

Finally, since $W_j \in \mathcal A_L = \operatorname{span}_{\mathbb C}\{P_v : v \in L\}$, its Pauli support is contained in $L$. Therefore $|\operatorname{supp}_P(W_j)| \le |L| = 2^r \le 4^m$.
\end{proof}

We now apply Theorem~\ref{thm:ham-learning-tomo-F} to learn $W_j$ in normalized Frobenius norm. 
Let $s := |\operatorname{supp}_P(W_j)| \le 4^m$, $c_F := 2\pi T \sqrt{m}$, and $c_\infty := 2\pi T m$. 
Choosing $c = \mathcal{O}(1/\sqrt{m})$, we obtain a total evolution time of
\begin{equation}
    \widetilde{\mathcal{O}}\!\left(\frac{s c_F c_\infty}{c^3 \varepsilon}\right)
= \widetilde{\mathcal{O}}\!\left(\frac{4^m T^2}{\varepsilon}\right)
\end{equation}
with respect to evolution under $W_j$.

We assume $\varepsilon \le 1/(m^2 T^2)$ and set $t = \mathcal{O}(1/(m^2 T^2 \varepsilon))$ as an integer. 
This enables queries to $C_j^q = e^{-iW_jq}$ for $q\in\mathbb{N}$. 
Since each query to $C_j^\dagger = e^{-iHT} e^{iH_jT}$ requires evolution time $T$ under $H$, the total evolution time with respect to $H$ is $\widetilde{\mathcal{O}}({4^m T^3}/{\varepsilon})$.

\subsection{Approximating continuous quantum control for many-body Hamiltonians}\label{subsec:manylong}

When $H$ is a many-body Hamiltonian, its sparsity scales as $m=\mathrm{poly}(n)$. 
From the previous section, we have $|\operatorname{supp}_P(W_j)| \le 4^m$, which becomes exponential in $n$. 
Thus, directly learning $W_j$ is infeasible.

To address this, we restrict the evolution time to scale with the sparsity, setting $T = \mathcal{O}(m^{-1/K})$ for some constant $K \in \mathbb{N}$. 
In this regime, we show that $W_j$ admits a quasi-sparse approximation via the BCH expansion.

\begin{lemma}[Quasi Pauli sparsity of the BCH generator]\label{prop:quasi-sparse-W}
Let $e^{iH_jT}e^{-iHT} = e^{-iW_j}$, where $H,H_j$ are $m$-sparse traceless Hermitian Hamiltonians satisfying 
$\|H\|_\infty,\|H_j\|_\infty \le C$ and $\|H-H_j\|_{\ell_\infty} \le \varepsilon$, 
for some constants $C \ge 1$, $\varepsilon > 0$, and $K \in \mathbb{N}$. Then for every $\ell \ge 1$, there exists $T = \Theta( m^{-1/K}/C)$ such that
\begin{equation}
    \operatorname{supp}_P\!\left(W_j,\frac{m^{-\ell/K} \varepsilon}{8} \right)
= \mathcal{O}(m^{\ell+K/2-1}),    
\end{equation}
and satisfies $\|W_j\|_F \le \sqrt{m}\varepsilon$ and $\|W_j\|_\infty \le m\varepsilon$.
\end{lemma}

\begin{proof}
Assuming $T \le 1/(8eC)$ satisfies the convergence condition of the Baker-Campbell-Hausdorff equation (Lemma~\ref{lem:BCH}), we have
\begin{equation}
    W_j = i\sum_{r\ge 1}\mathrm{BCH}_r(-iHT,iH_jT).
\end{equation}
Since $\|-iHT+iH_jT\|_F = T\|H-H_j\|_F \le T\sqrt{m}\varepsilon$ and $\max\{\|-iHT\|_\infty,\|iH_jT\|_\infty\} \le TC$, Corollary~\ref{cor:BCH_F_bound} yields
\begin{align}
\|\mathrm{BCH}_r(-iHT,iH_jT)\|_F &\le (4TeC)^r \sqrt{m}\,\varepsilon, \\
\|\mathrm{BCH}_r(-iHT,iH_jT)\|_\infty &\le (4TeC)^r m\varepsilon.
\end{align}
Moreover, $\operatorname{supp}_P(\mathrm{BCH}_r(-iHT,iH_jT)) \le (2m)^r$.

Define $W_j^{(k)} := i\sum_{r=1}^k \mathrm{BCH}_r(-iHT,iH_jT)$. Then
\begin{equation}
    \|W - W_j^{(k)}\|_F \le \sum_{r\ge k+1} (4TeC)^r \sqrt{m}\,\varepsilon \le (4TeC)^{k+1} \sqrt{m}\,\varepsilon.
\end{equation}
Furthermore,
\begin{equation}
    \operatorname{supp}_P(W_j^{(k)}) \le \sum_{r=1}^k (2m)^r \le k(2m)^k,
\end{equation}
and hence $\operatorname{supp}_P\left(W_j, (4TeC)^{k+1}\sqrt{m}\,\varepsilon\right) \le k(2m)^k$.

Choosing $T = {m^{-1/K}}/{(16eC)}$ and $k = {K}/{2} + \ell - 1$ yields
\begin{equation}
    \operatorname{supp}_P\!\left(W_j, \, \frac{m^{-\ell/K} \varepsilon}{8} \right)
= \mathcal{O}(m^{\ell+K/2-1}).    
\end{equation}

For this choice of $T$, we also have
\begin{align}
    \|W_j\|_F & \le \sum_{r\ge 1} (4TeC)^r \sqrt{m}\,\varepsilon \\
    & \le \sum_{r\ge 1} (1/2)^r \sqrt{m}\,\varepsilon\\
    & = \sqrt{m}\,\varepsilon,    
\end{align}
and similarly $\|W_j\|_\infty \le m\varepsilon$.
\end{proof}

We now aim to learn $W_j$ within error $c\varepsilon$ in normalized Frobenius norm. 
Suppose $c = \mathcal{O}(1/\sqrt{m})$. Setting $\ell = K/2$, we obtain 
$s = |\operatorname{supp}_P(W_j, c\varepsilon/8)| = \mathcal{O}(m^{K-1})$, 
and parameters $c_F = \sqrt{m}$ and $c_\infty = m$. 

Applying Theorem~\ref{thm:ham-learning-tomo-F}, we obtain a total evolution time of
\begin{equation}
    \widetilde{\mathcal{O}}\!\left(\frac{s c_\infty c_F T}{c^3 \varepsilon}\right)
= \widetilde{\mathcal{O}}\!\left(\frac{m^{K+2}T}{\varepsilon}\right).    
\end{equation}
Notably, Theorem~\ref{thm:ham-learning-tomo-F} remains applicable under this quasi-sparsity condition.

\section{The main algorithm}\label{sec:app-main}

In this section, we elaborate on the details of the main algorithm. We first define the global parameters $s, c_F, c_\infty$, where $s$ denotes the (quasi-)sparsity of $W_j$, and $c_F, c_\infty$ are constants such that
$\|W_j\|_F \le c_F \eta_j$ and $\|W_j\|_\infty \le c_\infty \eta_j$.
The desired error constant $c$ will be specified later. The values of these parameters are summarized in \Cref{alg:main}.

The algorithm follows a hierarchical refinement scheme that halves the error at each iteration:
\begin{equation}
    \|H - H_j\|_{\ell_\infty} \le \eta_j 
\;\;\Longrightarrow\;\;
    \|H - H_{j+1}\|_{\ell_\infty} \le \frac{\eta_j}{2} = \eta_{j+1}.
\end{equation}
At iteration $j$, we define the local parameters
\begin{equation}
    \eta_j = 2^{-j}, \qquad
    t_j = \frac{1}{32m\eta_j}, \qquad
    N_j = \frac{1}{2\sqrt{m}\eta_j}.
\end{equation}
After $J=\lceil \log_2(1/\varepsilon) \rceil$ iterations, the estimate $H_J$ satisfies $\|H - H_J\|_{\ell_\infty} \le \varepsilon$.

We now describe the $j$-th iteration in the regime $\eta_j \le \eta_{\mathrm{sw}}$.

\begin{enumerate}[(1)]

\item \textbf{Quantum control emulation (\Cref{alg:pure-Frobenius}).}

We define the emulated control unitary
\begin{equation}
\widetilde{U}_{\mathrm{res}}^{(j)}
:=
\Bigl(
e^{-iH(T+t_j/N_j)}\,e^{i\widetilde{W}_j}\,e^{iH_j(T+t_j/N_j)}
\Bigr)^{N_j},
\end{equation}
where $W_j$ satisfies $e^{-iW_j} = e^{iH_jT}e^{-iHT}$. In particular, for every $t \in \mathbb{N}$,
\begin{equation}
    e^{-iW_j t} = \bigl(e^{iH_jT}e^{-iHT}\bigr)^t,
\end{equation}
which can be implemented using only long-time evolutions.

Applying \Cref{alg:pure-Frobenius} with parameters $s, c_F, c_\infty$ and error constant $c = {1}/{(256\sqrt{m})}$, we obtain $\widetilde{W}_j$ such that
\begin{equation}
    \left\|W_j - \widetilde{W}_j \right\|_F \le c\eta_j = \frac{\eta_j}{256\sqrt{m}}.
\end{equation}

Consequently,
\begin{align} 
\left\|(e^{-iHt/N_j}e^{iH_jt/N_j})^{N_j}-\widehat U_{\mathrm{res}}^{(j)}\right\|_F &\le \left\|(e^{-iH\tau_j}e^{iW_j}e^{iH_j\tau_j})^{N_j}-(e^{-iH\tau_j}e^{i\widehat{W}_j}e^{iH_j\tau_j})^{N_j}\right\|_F \\ \nonumber 
&\le N_j \left\|e^{iW_j}-e^{i\widetilde{W}_j}\right\| \\ \nonumber 
&\le N_j \left\|W_j-\widetilde{W}_j \right\| \\ \nonumber &\le \frac{1}{512m}. 
\end{align}

By the Trotter formula,
\begin{equation}
\left\|(e^{-iHt_j/N_j}e^{iH_jt_j/N_j})^{N_j} - e^{-i\Delta H_j t_j} \right\|_F
\le \frac{t_j^2}{N_j}\sqrt{m}\eta_j \le \frac{1}{512m}.
\end{equation}

Therefore,
\begin{equation}
    \left\|\widetilde{U}_{\mathrm{res}}^{(j)} - e^{-i\Delta H_j t_j}\right\|_F \le \frac{1}{256m}.
\end{equation}

This $\mathcal{O}(1/m)$ control error is precisely the regime required for the next step.

The cost of learning $\widetilde{W}_j$ is
\begin{equation}
    \widetilde{\mathcal{O}}\!\left(\frac{4^m T^3}{\eta_j}\right)
\quad \text{or} \quad
    \widetilde{\mathcal{O}}\!\left(\frac{m^{K+2}T}{\eta_j}\right),
\end{equation}
for $m=\mathcal{O}(\log n)$ and $m=\mathrm{poly}(n)$, respectively.

\item \textbf{Coefficient and structure learning (\Cref{alg:pure-l-infty}).}

Using $\widetilde{U}_{\mathrm{res}}^{(j)}$, we approximate
\begin{equation}
    \ket{\psi_j}
    = V(e^{-i\Delta H_j t_j}\otimes I)\ket{\Phi}
    \simeq
    \ket{0} - it_j \sum_{x \neq 0} \alpha_x^{(j)} \ket{x}.
\end{equation}

Applying \Cref{alg:pure-l-infty}, we can learn $\Delta H_j$ up to error $\eta_j/8$ assuming access to $e^{-i\Delta H_j t_j}$. Since we instead use $\widetilde{U}_{\mathrm{res}}^{(j)}$, this introduces an additional error. Because
\begin{equation}
    \left\|\widetilde{U}_{\mathrm{res}}^{(j)} - e^{-i\Delta H_j t_j} \right\|_F \le \frac{1}{256m},    
\end{equation}
this contributes at most ${1}/{(256m t_j)} = {\eta_j}/{8}$ to the coefficient estimation error. Hence,
\begin{equation}
    \left\|\Delta H_j - \widetilde{\Delta H}_j\right\|_{\ell_\infty} \le \frac{\eta_j}{4}.
\end{equation}

This step uses $\mathcal{O}(m^3)$ queries to $\widetilde{U}_{\mathrm{res}}^{(j)}$. Since each query requires $N_j$ calls to $e^{-iH(T+t_j/N_j)}$, the total evolution time is
\begin{equation}
    \mathcal{O}(m^3 N_j T) = \mathcal{O}\!\left(\frac{m^{2.5}T}{\eta_j}\right).    
\end{equation}

\item \textbf{Sparse bounded truncation and update (\Cref{lem:sbt-stab}).}

We restore the sparsity and norm constraints by
\begin{equation}
    H_{j+1} := T_{m,1} \left(H_j + \widetilde{\Delta H}_j \right).
\end{equation}
By \Cref{lem:sbt-stab}, this ensures $H_{j+1} \in S_{m,1}$ and
\begin{equation}
\left\|H - H_{j+1} \right\|_{\ell_\infty}
\le
2 \left\|H - \left(H_j + \widetilde{\Delta H}_j\right) \right \|_{\ell_\infty}
\le
\frac{\eta_j}{2}
=
\eta_{j+1}.
\end{equation}

This step is purely classical and requires no additional quantum resources.

\end{enumerate}

The overall complexity is dominated by the cost of learning $\widetilde{W}_j$. Therefore,
\begin{equation}
    t_{\mathrm{tot}} = \sum_{j=1}^J \widetilde{\mathcal{O}}\!\left(\frac{4^m T^3}{\eta_j}\right)
    = \widetilde{\mathcal{O}}\!\left(\frac{4^m T^3}{\varepsilon}\right),
\end{equation}
for $m=\mathcal{O}(\log n)$, and
\begin{equation}
    t_{\mathrm{tot}} = \sum_{j=1}^J \widetilde{\mathcal{O}}\!\left(\frac{m^{K+2}T}{\eta_j}\right)
    = \widetilde{\mathcal{O}}\!\left(\frac{m^{K+2}T}{\varepsilon}\right),
\end{equation}
for $m=\mathrm{poly}(n)$ with $T=\Theta(m^{-1/K})$, where $K\in\mathbb{N}$.

\section{Hamiltonian learning with long time dynamics in the large error regime}\label{sec:sql}

\begin{algorithm}
\caption{$\textsc{LearnStandardQuantumLimit}(\mathcal{U}_A,A_0,m,\varepsilon)$}
\label{alg:sql}
\DontPrintSemicolon

\KwIn{Query access to $\mathcal{U}_A(\tau)=e^{-iA\tau}$ for an $m$-sparse traceless Hermitian operator $A$, and a known $m$-sparse Hamiltonian $A_0$ satisfying $\|A-A_0\|_{\ell_\infty}\le \varepsilon$ with $\operatorname{supp}_P(A)=\operatorname{supp}_P(A_0)=m$.}
\KwOut{An estimate $\widetilde A$ such that $\|A-A_0-\widetilde A\|_{\ell_\infty}\le \varepsilon/4$.}

Set $t \gets 1/(16\sqrt{m})$ and $\delta_t \gets \sqrt{m}\varepsilon t^2$\;

Prepare the maximally entangled state $\ket{\Phi}$ and a unitary $V$ such that $V(P_x\otimes I)\ket{\Phi}=\ket{x}$ for all $x$\;

Define $\mathcal{U}_{A,A_0}(T) \gets \mathcal{U}_A(T)e^{iA_0T}$ and $\mathcal{U}_{A,A_0}(T+t) \gets \mathcal{U}_A(T+t)e^{iA_0(T+t)}$\;

Prepare $\ket{\psi_T} \gets V(\mathcal{U}_{A,A_0}(T)\otimes I)\ket{\Phi}$ and $\ket{\psi_{T+t}} \gets V(\mathcal{U}_{A,A_0}(T+t)\otimes I)\ket{\Phi}$\;

\If{$m=\mathcal{O}(\log n)$}{
    $s \gets 4^m$\;
}
\ElseIf{$m=\mathrm{poly}(n)$ and $T=\Theta(m^{-1/K})$ for some $K\in\mathbb{N}$}{
        $s \gets \mathcal{O}(m^{K-1})$\;
}

Apply sparse pure-state tomography to learn $\ket{\psi_T}$ and $\ket{\psi_{T+t}}$ to accuracy $\delta_t$ using $\mathcal{O}(s/\delta_t^2)$ copies, obtaining coefficients $\{\widetilde{\alpha}_x^{(T)}\}_x$ and $\{\widetilde{\alpha}_x^{(T+t)}\}_x$ satisfying
\begin{equation}
    \left(\min_{\phi}\sum_x \left|\alpha_x^{(T)}-e^{i\phi}\widetilde{\alpha}_x^{(T)}\right|^2\right)^{1/2}\le \delta_t/2,
\quad
\left(\min_{\phi}\sum_x \left|\alpha_x^{(T+t)}-e^{i\phi}\widetilde{\alpha}_x^{(T+t)}\right|^2\right)^{1/2}\le \delta_t/2.    
\end{equation}

Form $\widetilde{\mathcal{U}}_{A,A_0}(T) \gets \sum_x \widetilde{\alpha}_x^{(T)} P_x$ and $\widetilde{\mathcal{U}}_{A,A_0}(T+t) \gets \sum_x \widetilde{\alpha}_x^{(T+t)} P_x$\;

Compute $\widetilde{\mathcal{U}}_{A,A_0}(t) \gets e^{iA_0T}\widetilde{\mathcal{U}}_{A,A_0}(T)^\dagger \widetilde{\mathcal{U}}_{A,A_0}(T+t)e^{-iA_0T}$\;

Let $\widetilde{\alpha}_0$ be the identity Pauli coefficient of $\widetilde{\mathcal{U}}_{A,A_0}(t)$\;

\Return{$\widetilde A \gets -\operatorname{Im}(\widetilde{\alpha}_0^{*}\widetilde{\mathcal{U}}_{A,A_0}(t))/t$}\;
\end{algorithm}

Algorithm~\ref{alg:sql} reconstructs the classical representation (Pauli coefficients) of the unitary $\mathcal{U}_{A,A_0}(t)=e^{-iAt}e^{iA_0t}$. 
Assuming $\|A-A_0\|_{\ell_\infty} \le \varepsilon$, for $t \le 1/\sqrt{m}$ we have
\begin{align}
    \left\|\mathcal{U}_{A,A_0}(t)-I+i(A-A_0)t\right\|_{\ell_\infty} & \le \|\mathcal{U}_{A,A_0}(t)-I+i(A-A_0)t\|_{F} \\
    & \le 
    \sqrt{m}\varepsilon t^2 = \delta_t.    
\end{align}

Suppose we obtain $\widetilde{\mathcal{U}}_{A,A_0}(t)$ satisfying $\min_{\phi}\|\widetilde{\mathcal{U}}_{A,A_0}(t)-e^{i\phi}\mathcal{U}_{A,A_0}(t)\|_F \le \delta_t$, and write $\mathcal{U}_{A,A_0}(t)=\sum_x \alpha_x P_x$ and $\widetilde{\mathcal{U}}_{A,A_0}(t)=\sum_x \widetilde{\alpha}_x P_x$. 
Then $|e^{i\phi}\widetilde{\alpha}_x-\alpha_x| \le \delta_t$ and $|1-\alpha_0| \le \delta_t$, which implies
\begin{equation}
    \left|e^{i\phi}-\widetilde{\alpha}_0\right| \le |1-\alpha_0| + \left|e^{i\phi}\alpha_0 - \widetilde{\alpha}_0\right| \le 2\delta_t.    
\end{equation}
Hence $|\alpha_x - \widetilde{\alpha}_0^*\widetilde{\alpha}_x| \le 3\delta_t$ for all $x$, and thus $\|\widetilde{\alpha}_0^*\widetilde{\mathcal{U}}_{A,A_0}(t)-I+i(A-A_0)t\|_{\ell_\infty}
\le 4\delta_t$.

Finally,
\begin{equation}
    \left\|\widetilde A - (A-A_0)\right\|_{\ell_\infty} 
    \le \frac{1}{t}\left\|\widetilde{\alpha}_0^*\widetilde{\mathcal{U}}_{A,A_0}(t)-I+i(A-A_0)t\right\|_{\ell_\infty}
    \le \frac{4\delta_t}{t}
    \le \frac{\varepsilon}{4},    
\end{equation}
where $\widetilde A = -\operatorname{Im}(\widetilde{\alpha}_0^*\widetilde{\mathcal{U}}_{A,A_0}(t))/t$.

We now show how to reconstruct $\mathcal{U}_{A,A_0}(t)$ using only access to $\mathcal{U}_A(\tau)=e^{-iA\tau}$ for $\tau \ge T$. 
Observe that
\begin{equation}
    \mathcal{U}_{A,A_0}(t)
=
e^{iA_0T}\mathcal{U}_{A,A_0}(T)^\dagger \mathcal{U}_{A,A_0}(T+t)e^{-iA_0T}.    
\end{equation}
Thus, learning the Pauli coefficients of $\mathcal{U}_{A,A_0}(T)$ and $\mathcal{U}_{A,A_0}(T+t)$ suffices.

From Section~\cref{sec:conquantumlong}, we define
$s_T := \operatorname{supp}_P(\mathcal{U}_{A,A_0}(T))$
and
$s_{T+t} := \operatorname{supp}_P(\mathcal{U}_{A,A_0}(T+t))$.

When $m=\mathcal{O}(\log n)$, we have $s_T, s_{T+t} \le 4^m$.

When $m=\mathrm{poly}(n)$ and $T=\mathcal{O}(m^{-1/N})$, we instead define the quasi-sparsities
$s_T := \operatorname{supp}_P(\mathcal{U}_{A,A_0}(T),\delta_t)$
and
$s_{T+t} := \operatorname{supp}_P(\mathcal{U}_{A,A_0}(T+t),\delta_t)$,
for which $s_T, s_{T+t} = \mathcal{O}(m^{K-1})$.

Applying sparse pure-state tomography (Lemma~\ref{lem:sparse-tomo-l2}), we obtain estimates $\widetilde{\mathcal{U}}_{A,A_0}(T)$ and $\widetilde{\mathcal{U}}_{A,A_0}(T+t)$ satisfying $d(\mathcal{U}_{A,A_0}(T),\widetilde{\mathcal{U}}_{A,A_0}(T)) \le \delta_t/2$ and $d(\mathcal{U}_{A,A_0}(T+t),\widetilde{\mathcal{U}}_{A,A_0}(T+t)) \le \delta_t/2$ using $\widetilde{\mathcal{O}}(\max\{s_T,s_{T+t}\}/\delta_t^2)$ queries.

This yields $d(\widetilde{\mathcal{U}}_{A,A_0}(t), \mathcal{U}_{A,A_0}(t)) \le \delta_t$, where $\widetilde{\mathcal{U}}_{A,A_0}(t) = e^{iA_0T}\widetilde{\mathcal{U}}_{A,A_0}(T)^\dagger \widetilde{\mathcal{U}}_{A,A_0}(T+t)e^{-iA_0T}$.

Each query to $\mathcal{U}_{A,A_0}(T)$ or $\mathcal{U}_{A,A_0}(T+t)$ requires time $T$. 
Hence, the total evolution time is 
$\widetilde{\mathcal{O}}(4^m T/\varepsilon^2)$ or 
$\widetilde{\mathcal{O}}(m^N T/\varepsilon^2)$, 
while the classical post-processing time is $\mathrm{poly}(s)/\varepsilon^2$, 
where $s := \max\{s_T, s_{T+t}\}$.
\end{document}